\documentclass[12pt]{article}
\usepackage{latexsym}
\usepackage[bf,footnotesize]{caption}	
\begin{document}
\setlength{\unitlength}{1mm}

\newcommand{\be}{\begin{equation}}
\newcommand{\ee}{\end{equation}}
\newcommand{\ben}{$$}
\newcommand{\een}{$$}
\newcommand{\bea}{\begin{eqnarray}}
\newcommand{\eea}{\end{eqnarray}}
\newcommand{\bean}{\begin{eqnarray*}}
\newcommand{\eean}{\end{eqnarray*}}
\newcommand{\e}{{\rm e}}
\newcommand{\tr}{{\rm tr}}

\newcommand{\n}[1]{\label{#1}}
\newcommand{\eq}[1]{Eq.(\ref{#1})}
\newcommand{\ind}[1]{\mbox{\tiny{#1}}}
\renewcommand\theequation{\thesection.\arabic{equation}}

\newcommand{\nn}{\nonumber \\ \nonumber \\}
\newcommand{\nl}{\\  \nonumber \\}
\newcommand{\pr}{\partial}
\renewcommand{\vec}[1]{\mbox{\boldmath$#1$}}

\title{{\hfill {\small Alberta-Thy-15-99 } } \vspace*{2cm} \\
The Dimensional-Reduction Anomaly}
\author{\\
V. Frolov\thanks{e-mail: frolov@phys.ualberta.ca} 
\ and \ P. Sutton\thanks{e-mail: psutton@phys.ualberta.ca} \\ 
{\em \small Theoretical Physics Institute, Department of Physics,} \\
{\em \small University of Alberta, Edmonton, Canada T6G 2J1}  \\
\\ A. Zelnikov\thanks{e-mail: zelnikov@phys.ualberta.ca}  \\
{\em \small Theoretical Physics Institute, Department of Physics,} \\
{\em \small University of Alberta, Edmonton, Canada T6G 2J1}      \\
{\em \small and Lebedev Physics Institute, Leninsky Prospect 53,} \\
{\em \small Moscow 117924, Russia}} 
\date{Oct. 9, 1999}
\maketitle
\noindent

\begin{abstract}
In a wide class of $D$-dimensional spacetimes which are direct or
semi-direct sums of a $(D-n)$-dimensional space and an $n$-dimensional 
homogeneous ``internal'' space, a field can be decomposed into modes. 
As a result of this mode decomposition, the main objects which
characterize the free quantum field, such as Green functions and heat
kernels, can effectively be reduced to objects in a $(D-n)$-dimensional
spacetime with an external dilaton field.  We study the problem of the 
dimensional reduction of the effective action for such spacetimes. 
While before renormalization the original $D$-dimensional effective
action can be presented as a ``sum over modes'' of $(D-n)$-dimensional
effective actions, this property is violated after renormalization.  We
calculate the corresponding anomalous terms 
explicitly, illustrating the effect  with some simple examples.
\end{abstract}
\vspace{.3cm}


\section{Introduction} 
Simplifications connected with an assumption of symmetry play an
important role in the study of physical effects in a curved spacetime.
In this paper we consider quantum fields propagating in a
$D$-dimensional spacetime which is a semi-direct sum  of a
$(D-n)$-dimensional space and an $n$-dimensional  homogeneous
``internal'' space. Such a field can be decomposed into modes.  As a
result of this mode decomposition, the main objects which characterize
the free quantum field, such as Green functions and heat kernels, can
effectively be reduced to objects in a $(D-n)$-dimensional spacetime
with an external dilaton field.  Our aim is to study the problem of
the  dimensional reduction of the effective action for such
spacetimes.  We shall demonstrate that while before renormalization the
original $D$-dimensional effective action can be presented as a ``sum
over modes'' of $(D-n)$-dimensional effective actions, this property is
generically violated after renormalization. We call this effect the {\em
dimensional-reduction anomaly}.

First of all, there is an evident general reason why $D$-dimensional 
renormalization is not equivalent to renormalization of 
the $(D-n)$-dimensional effective theory.  Namely, the number of
divergent terms of the Schwinger-DeWitt series which is used to
renormalize a given object depends on the number of dimensions. What is
much more interesting is that this is not the only reason for the
presence of the dimensional-reduction anomaly. The aim of this paper is
to discuss this problem and to derive explicit expressions for the 
dimensional-reduction anomaly in a four-dimensional spacetime with 
special choices of one- and two-dimensional homogeneous internal spaces.

In some aspects the problem we study is related to the study of the
momentum-space representation of the ultraviolet divergences discussed
by Bunch and Parker \cite{BuPa:79}. Nevertheless, there exists a very
important difference. Namely, Bunch and Parker used the Fourier
transform with respect to all $D$ dimensions ($D=4$ in their paper) in a
spacetime without symmetries. We are making mode decompositions with
respect to $n$ internal dimensions only. Moreover, due to the symmetry we
effectively rewrite the original $D$-dimensional theory in terms of a
set of $(D-n)$-dimensional effective theories with a dilaton field.  
This representation, which was absent in the paper by Bunch and Parker,
allows us to make the comparison of renormalization in $D$- and
$(D-n)$-dimensional theories.

The dimensional-reduction anomaly discussed in this paper might have 
interesting applications. One of them is connected with black-hole
physics. Recently there has been much interest in the study of the
Hawking radiation in two-dimensional dilaton gravity models of a black
hole. This study was initiated by the paper \cite{MuWiZe:94}. In this
and other papers on the subject (see e.g.
\cite{ChSi:97}--\cite{BaFa:99}) it is either explicitly or implicitly
assumed that two-dimensional calculations (at least for the special choice
of the dilaton field corresponding to the spherical reduction of the
four-dimensional Schwarzschild spacetime) correctly reproduce the $s$-mode
contribution to the stress-energy tensor of the four-dimensional
theory. Generally speaking, in the presence of the 
dimensional-reduction anomaly this is not true (see also the discussion
in \cite{NoOd:99}).

It is interesting to note that in a numerical study of the vacuum
polarization in black holes where the mode decomposition was used, it has
been demonstrated that one always needs to add extra contributions to the
terms of the series in order to ensure convergence of the series
(see e.g. \cite{Cand:80,Ande:90,AnHiSa:95}). This fact, as we shall see,
is a direct manifestation of the dimensional-reduction anomaly.

The paper is organized as follows.  Section~\ref{s2} contains a general
discussion of the dimensional reduction of a free quantum field theory
in a gravitational background. In Section~\ref{s3} we discuss simple examples
of the dimensional reduction of four-dimensional flat spacetime and
illustrate the effect of the dimensional-reduction anomaly. In
Sections~\ref{s4} and \ref{s5} we derive the dimensional-reduction anomaly for
$\langle \hat{\Phi}^2\rangle^{\ind{ren}}$ and $\langle
\hat{T}^{\mu}_{\nu}\rangle^{\ind{ren}}$ in a four-dimensional spacetime
for $(3+1)$ and $(2+2)$ reductions. We discuss the obtained results in
the Conclusion. In our work we use dimensionless units where 
$G=c=\hbar=1$, and the sign conventions of \cite{MTW} for the 
definition of the curvature.


\section{Dimensional Reduction of the Heat Kernel and the Effective 
Action}\label{s2}
\setcounter{equation}0
Consider a $D$-dimensional spacetime with a metric of the form
\be\n{2.1}
ds^2  =  g_{\mu\nu}\, dX^{\mu}\, dX^{\nu}
      =  dh^2 +e^{-(4\phi/n)} d\Omega^2\, ,  \qquad  X^{\mu} =(x^a, y^i)\, ,
\ee
\be\n{2.2}
\phi  =  \phi(x^c)\, , \qquad
dh^2  =  h_{ab}(x^c) dx^adx^b\, ,\qquad
d\Omega^2  =  \Omega_{ij}(y^k) dy^i dy^j\, ,
\ee
where $d\Omega^2$ is the metric of an $n$-dimensional
homogeneous space. In other words, the metric $ds^2$ is a semi-direct sum of
$dh^2$ and $d\Omega^2$. We call the $n$-dimensional space with 
metric $d\Omega^2$ the {\em internal space}, and the scalar field
$\phi(x^c)$ on the $(D-n)$-dimensional manifold with metric $dh^2$
the {\em dilaton field}. Let us emphasize that the normalization of the
dilaton field $\phi$ is a question of convenience. We fix this
normalization by requiring that $\sqrt{g}$ for the metric (\ref{2.1}) be
proportional to $\exp(-2\phi)$ for any number of internal dimensions $n$.
Well-known examples of metrics of the form (\ref{2.1}) are those of 
spherical spacetimes, and metrics connected with a dimensional reduction 
in Kaluza-Klein theories. 

Let $\hat{\Phi}$ be a free scalar quantum field propagating in the spacetime
(\ref{2.1}) and obeying the equation
\be\n{2.3}
F \, \hat{\Phi}(X) = 0 \, ,
\ee
with field operator
\be\n{2.3b}
F  =  \Box-V-m^2 \, .
\ee
Note that we explicitly separate the mass term $m^2$ from the potential $V$.  
The latter may contain an interaction with the curvature, $\xi R$, for a
non-minimally coupled field, but is not fixed at the moment. We only
assume that when calculated on the background (\ref{2.1}) the potential
$V$ is independent of the $y^a$ coordinates.

Using the line element (\ref{2.1}), the operator $\Box$ becomes
\be\n{2.4}
\Box  =  \Delta_h - 2 \nabla\phi\cdot\nabla 
         + {\rm e}^{(4\phi/n)}\Delta_\Omega\, , 
\ee
where $\Delta_h$, $\Delta_\Omega$ are the  d'Alembertians
corresponding to the metrics $h_{ab}$, $\Omega_{ij}$ respectively, and
$\nabla$ is understood to denote the covariant derivative with respect to the
metric $h_{ab}$.

Considerable simplification of the problem in spacetime (\ref{2.1}) is
connected with the fact that for a wide class of homogeneous metrics
the eigenvalue problem
\be\n{2.5}
\Delta_\Omega Y(y)= -\lambda Y(y)\, , 
\ee
is well-studied. We denote by $Y_{\lambda W}$ harmonics, that is  
eigenfunctions of (\ref{2.5}), and use a (collective)  index $W$ to
distinguish between different solutions of (\ref{2.5}) for the same
$\lambda$. We assume standard normalization and orthogonality
conditions,\footnote{We write summation over indices assuming that the
spectrum is discrete. For a continuous spectrum one must replace 
summation by integration over the spectrum. 
In what follows we shall assume that this rule is automatically applied.}
\be\n{2.6}
\int dy\,  |\Omega|^{1/2} \, Y_{\lambda W} (y) \bar{Y}_{\lambda' W'}(y)
  =  \delta_{\lambda \lambda'}\delta_{W W'}\, ,
\ee
\be\n{2.7}
\sum_{\lambda W} Y_{\lambda W} (y) \bar{Y}_{\lambda W}(y')
  =  \delta(y,y')\equiv |\Omega|^{-1/2} \delta^{n}(y-y') \, .
\ee
We denote by ${\cal N}(\lambda)$ the degeneracy of the eigenvalue $\lambda$,
that is ${\cal N}(\lambda)=\sum_{W,W'}\delta_{W W'}$.

The field $\hat{\Phi}$ can be decomposed into {\em modes} 
\be\n{2.8}
e^{\phi} \varphi_{\lambda p}(x) Y_{\lambda W}(y)\, ,
\ee
where the functions $\varphi_{\lambda p}$ obey the equation
\be\n{2.9}
{\cal F}_{\lambda} \varphi_{\lambda p}(x) = 0\, ,
\ee
\be\n{2.10}
{\cal F}_{\lambda} =\Delta_h - V_{\lambda}[\phi] - m^2\, .
\ee
The index $p$ is an additional quantum number which enumerates 
solutions for a given $\lambda$ and $V_{\lambda}[\phi]$ is
\be\n{2.11}
V_{\lambda}[\phi]  
  =  \lambda {\rm e}^{(4\phi/n)} + (\nabla\phi)^2 - \Delta_h\phi  + V\, .
\ee
In other words, by expanding the field in modes we effectively reduce
the original $D$-dimensional problem to a similar problem in
$(D-n)$-dimensional space with an effective potential $V_\lambda$ 
depending on the ``dilaton'' field $\phi$.

It is not difficult to show that at least formally the effective action
for the quantum field (\ref{2.3}) allows a similar dimensional reduction.
In order to demonstrate this, consider a {\em heat kernel} $K(X,X'|s)$
for the problem (\ref{2.3}), which is the solution of the equation 
\be\n{2.12}
\left( {\partial\over \partial s} - F \right)K(X,X'|s) = 0 \, ,
\ee
\be\n{2.13}
K(X,X'|s=0)
  =  \delta^D(X,X') 
  =  e^{2\phi(x)} {\delta^{(D-n)}(x-x')\over\sqrt{h}} 
     {\delta^n(y-y')\over \sqrt{\Omega}}\, .
\ee
The effective action is defined as
\be\n{2.14}
W = -\frac{1}{2} \int_0^\infty \frac{ds}{s} tr_X K(X,X'|s)\, .
\ee
Here and later the trace operation is understood as
\be\n{2.15}
tr_X A(X,X')
  =  \int \, dX \sqrt{g}\, A(X,X)
  =  \int \, dx \sqrt{h} \int \, dy \sqrt{\Omega}\,  
     e^{-2\phi(x)}\, A(x,y;x,y) \, .
\ee
By decomposing $K(X,X'|s)$ into harmonics one can write
\be\n{2.16}
K(X,X'|s) 
  =  e^{\phi(x)+\phi(x')} \, \sum_{\lambda, W}
     {\cal K}(x,x';\lambda|s)\, Y_{\lambda W} (y) \bar{Y}_{\lambda W}(y')\, .
\ee
Using relations (\ref{2.4})--(\ref{2.7}), it is easy to verify that 
the {\em reduced heat kernel} ${\cal K}(x,x';\lambda|s)$ obeys the relations
\be\n{2.17}
\left( {\partial\over \partial s} - {\cal F}_{\lambda}\right) 
{\cal K}(x,x';\lambda|s) = 0\, ,
\ee
\be\n{2.18}
{\cal K}(x,x';\lambda|s=0) 
  =  \delta^{(D-n)}(x,x') 
  =  {\delta^{(D-n)}(x-x')\over\sqrt{h}}\, .
\ee
Here the operator ${\cal F}_{\lambda}$ is given by (\ref{2.9})--(\ref{2.10}).

Substituting representation (\ref{2.16}) into the effective action
(\ref{2.14}), integrating over the $y$-variables, and using (\ref{2.6}),  
one gets
\be\n{2.19}
W = \sum_{\lambda} {\cal N}(\lambda) W_{\lambda}\, ,
\ee
where
\be\n{2.20}
W_{\lambda}= -\frac{1}{2}  \int_0^\infty \frac{ds}{s} tr_x 
{\cal K}(x,x';\lambda|s)\, .
\ee
Here
\be\n{2.21}
tr_x  {\cal K}(x,x';\lambda|s)
  =  \int dx \, \sqrt{h} \, {\cal K}(x,x;\lambda|s)\, ,
\ee
and ${\cal N}(\lambda)$ is the degeneracy factor of the eigenvalue $\lambda$. 
Relations (\ref{2.19})--(\ref{2.20}) can be interpreted
as the mode decomposition of the effective action. 

It should be emphasized that the above relations for the effective
action are strictly {\em formal} because of the presence of ultraviolet
divergences. In order to obtain the renormalized value of the effective
action in $D$-dimensional spacetime one must subtract from $tr_X
K(X,X'|s)$ the first $N_D$ terms of the 
Schwinger-DeWitt expansion of the heat kernel, where
\be
N_D = \left\{ \begin{array}{c c} 
            \frac{D}{2}+1   &  {\rm for}\ \  $D$ {\rm \; even ,} \vspace{0.3cm} \\
            \frac{D+1}{2} & {\rm for}\ \  $D$ {\rm \; odd .}
            \end{array} \right. 
\ee
Our main observation is that this procedure destroys the formal
representation (\ref{2.19}), so that after renormalization one gets
\be\n{2.22}
W^{\ind{ren}} = \sum_{\lambda} {\cal N}(\lambda) 
\left[W^{\ind{ren}}_{\lambda}+\Delta
W_{\lambda}\right]\, .
\ee
In this expression $W^{\ind{ren}}_{\lambda}$ is understood as the
renormalized effective action of the $(D-n)$-dimensional theory with
a dilaton field, (\ref{2.9}), where the renormalization is performed by
subtracting the first $N_{D-n}$ terms of the Schwinger-DeWitt expansion 
for the operator ${\cal F}_{\lambda}$. We call the additional contribution 
$\Delta W_{\lambda}$ the {\em dimensional-reduction anomaly}. 
A representation similar to (\ref{2.22}) is also valid for 
$\langle \hat{\Phi}^2\rangle^{\ind{ren}}$,
\be\n{2.23}
\langle \hat{\Phi}^2\rangle^{\ind{ren}}  
  =  \e^{2\phi} \sum_{\lambda} {\cal N}(\lambda) 
     \left[ \langle \hat{\Phi}^2\rangle^{\ind{ren}}_{\lambda}+\Delta
         \langle \hat{\Phi}^2\rangle_{\lambda} \right] \, .
\ee

One might observe that there exists a relationship between 
the dimensional-reduction anomaly and the so-called {\em multiplicative 
anomaly} \cite{ElVaZe:98,Do:98}. Formally, we can write
\be\n{2.24}
F  =  \prod_{\lambda, W} {\cal F}_{\lambda}\, ,
\ee
\be\n{2.25}
-{1\over 2}\log \det F  
  =  -{1\over 2}\sum_{\lambda} N(\lambda) \log \det {\cal F}_{\lambda}\, .
\ee
The latter relation is nothing but (\ref{2.19}). The
violation of the formal relation (\ref{2.25}) for products of operators
after renormalization is known as the multiplicative anomaly.

The aim of this paper is to discuss special examples of the
dimensional-reduction anomaly. In what follows, we restrict ourselves to the
physically interesting case where the number of spacetime dimensions is
4, and the number of dimensions of the ``internal'' homogeneous space is
1 or 2.  We also restrict ourselves to manifolds of Euclidean signature.  
In each case the anomaly is found as the difference between the 
renormalization terms for the $(D-n)$-dimensional theory 
and the mode-decomposed renormalization terms from $D$ dimensions.


\section{Flat Space Examples of the Dimensional-Reduction Anomaly}\label{s3}
\setcounter{equation}0


The dimensional-reduction anomaly can occur in even the simple case
of mode decomposition in a flat spacetime.  In order to demonstrate 
this, let us consider a free massive scalar field $\hat{\Phi}$ obeying 
the equation
\be\n{3.1}
F\, \hat{\Phi}= (\Box -m^2) \hat{\Phi}=0\, 
\ee
in four-dimensional flat Euclidean space. The Euclidean Green function for
this equation is
\be\n{3.2}
G^0(X,X')={m\over 4\pi^2\sqrt{2\sigma}} K_1(m\sqrt{2\sigma})\, ,
\ee
where $2\sigma$ is the square of the geodesic distance from $X$ to
$X'$, and $K_1$ is a modified Bessel function. 
If there exists a boundary $\Sigma$ surrounding the region ${\cal M}$ 
under consideration and the field obeys a non-trivial boundary
condition at $\Sigma$, or equation (\ref{3.1}) includes a non-vanishing
potential $V$ which vanishes in ${\cal M}$, then the Green
function would differ from $G^0$. We denote this Green function by 
$G(X,X')$. The renormalized Green function in this case is defined 
as\footnote{In a generic four-dimensional curved background, 
one has to subtract the first two terms in the Schwinger-DeWitt 
expansion for the heat kernel of the operator $F$.}\label{fourdim}
\be\n{3.3}
G^{\ind{ren}}(X,X')=G(X,X')-G^0(X,X')\, .
\ee
We also have
\be\n{3.4}
\langle \hat{\Phi}^2(X)\rangle^{\ind{ren}}= \lim_{X'\rightarrow X} \,
G^{\ind{ren}}(X,X') \, .
\ee
It is evident that $\langle \hat{\Phi}^2(X)\rangle^{\ind{ren}}=0$  in
the absence of the external field $V$ and boundaries. 

In this section we derive the dimensional-reduction anomaly 
in $\langle \hat{\Phi}^2\rangle^{\ind{ren}}$ for two examples
of mode decompositions in flat space.  In each case, the anomaly is calculated
as the difference between the subtraction terms used to renormalize 
the Green function ${\cal G}$ of the dimensionally-reduced theory, and the
$G^0$ used to renormalize the four-dimensional Green function $G$.  
These calculations are easily repeated for the heat kernel, allowing
one to obtain the anomaly in the effective action in a similar manner.

\subsection{Spherical reduction}
\label{s3A}

For the first example we consider the case when the external field and/or
boundary is spherically symmetric and perform decomposition into
spherical harmonics.  The metric in spherical coordinates
$X^{\mu}=(t,r,\theta,\varphi)$ is given by
\be\n{3.5}
ds^2  =  dt^2 + dr^2 + r^2 \left( d\theta^2 + \sin^2\theta d\varphi^2 \right)  
         \, ,
\ee
and for the square of the geodesic distance from $X$ to $X'$ we have
\be\n{3.6}
2 \sigma = 
(\Delta t)^2 + (\Delta r)^2 + 2rr'\left(1 - \cos{\lambda} \right)\, ,
\ee
where $\Delta t =t-t'$, $\Delta r=r-r'$, and
\be\n{3.7}
\cos\lambda = \cos\theta\, \cos\theta'+\sin\theta\, \sin\theta'\,
\sin(\varphi -\varphi')\, .
\ee
The Green function $G^0$ can be decomposed into spherical harmonics
$Y_{\ell m}(\theta,\varphi)$,
\be\n{3.8}
G^0(X,X')  =  \sum_{\ell=0}^{\infty} \sum_{m=-\ell}^{\ell} 
              Y_{\ell m}(\theta,\varphi) \bar{Y}_{\ell m}(\theta',\varphi') \, 
              {{\cal G}^0_{\ell}(x,x')\over r\, r'}
           =  \frac{1}{4\pi} \sum_{\ell=0}^{\infty} (2\ell +1) \, 
              P_{\ell}(\cos{\lambda})\, 
              {{\cal G}^0_{\ell}(x,x')\over r\, r'} \, ,
\ee
where $x=(t,r)$ and $x'=(t',r')$. The two-dimensional Green function is
a solution of the equation
\be\n{3.9}
{\cal F}_{\ell}\, \, {\cal G}^0_{\ell}(x,x')\equiv 
({}^2\Box-m^2 -{V}_{\ell}){\cal G}^0_{\ell}(x,x')= -\delta(t-t')\,
\delta(r-r')\, ,
\ee
where ${}^2\Box =\partial_t^2 +\partial_r^2$, and
\be\n{3.10}
{V}_{\ell}(r)  =  {\ell(\ell +1) \over r^2}\, .
\ee

It is possible to write representations similar to (\ref{3.8}) for
the Green functions $G$ and $G^{\ind{ren}}$, and using these
representations to write the renormalized value of $\langle\hat{\Phi}^2\rangle$ in the form
\be\n{3.11}
\langle \hat{\Phi}^2(x)\rangle^{F}_{\ind{ren}}
  =  {1\over 4\pi r^2}\lim_{x'\rightarrow x} 
     \sum_{\ell=0}^{\infty} (2\ell +1) \, 
     {\cal G}^{\ind{ren}}_{\ell}(x,x') \, ,
\ee
where
\be\n{3.12}
{\cal G}^{\ind{ren}}_{\ell}(x,x')
  =  {\cal G}_{\ell}(x,x')-{\cal G}^{0}_{\ell}(x,x')\, .
\ee

The Green functions ${\cal G}^0_{\ell}$ can be obtained either by solving
equation (\ref{3.9}) or by decomposing the known function $G^0$ into
spherical harmonics. Decomposing $G^0$ we get
\be\n{3.13}
{{\cal G}^0_{\ell}(x,x')\over r\, r'} 
   =    {1\over 2\pi} \int_{-1}^1 dz~P_l(z)~
	{mK_1(m\sqrt{(\Delta t)^2+(\Delta r)^2+2 r r'(1-z)})
	\over\sqrt{(\Delta t)^2+(\Delta r)^2+2 r r'(1-z)}}
	\, ,
\ee
with $z=\cos\lambda$.
Using the integral representation for $K_1$ (see Appendix~\ref{K_formulae}) 
this becomes 
\be\n{3.14}
{{\cal G}^0_{\ell}(x,x')\over r\, r'}
   =   {1\over 8\pi} \int_0^\infty {ds \over s^2}~
	\exp\left( -m^2 s-{(\Delta t)^2+(\Delta r)^2+2 r r'\over 4 s} \right)
	\int_{-1}^1 dz~P_{\ell}(z)~
	\exp\left( {2 r r'\over 4 s}z \right)\, .
\ee
The integral over $z$ can be taken (see e.g. \cite{PrBrMa:86}, vol.2,
eq.2.17.5.2):
\be\n{3.15}
\int_{-1}^1 dz~P_{\ell}(z)~
	\exp\left(-p z\right)=(-1)^{\ell} \sqrt{2\pi\over p}I_{\ell+1/2}(p)\, ,
\ee
where $I_{\ell+1/2}$ is a modified Bessel function. Since for $\ell\ge 1$ the
expression in the right-hand side of (\ref{3.15}) vanishes at $p=0$, 
one can easily show that the functions ${\cal G}^0_{\ell}(x,x')$ vanish 
for all $\ell$ when either $r=0$ or $r'=0$.  This property will be of some
importance for the anomaly.

Using the following representation for the function $I_{\ell+1/2}$
\be\n{3.16}
I_{\ell+1/2}(p)
  =  {1\over\sqrt{2\pi p}}
     \sum_{k=0}^\ell {(\ell+k)!\over k!(\ell-k)!}~{1\over (2 p)^k}
     \left[ (-1)^k \e^p - (-1)^\ell e^{-p} \right] \, ,
\ee
(see, for example, 8.467 of \cite{GrRy:94}),
we obtain for the reduced Green functions
\bea
{\cal G}^0_{\ell}(x,x')
  & = &  {1\over 2\pi} 
         \sum_{k=0}^{\ell}{(\ell+k)!\over k!(\ell-k)!} \left[
         (-1)^{k}~{( (\Delta t)^2+(\Delta r)^2)^{k/2}\over (2m r r' )^k}
         ~K_k(m\sqrt{(\Delta t)^2+(\Delta r)^2})   \right. \nonumber \\
  &   &  \hspace{-0.15in} \left. \mbox{}
         -(-1)^{\ell}
         ~{( (\Delta t)^2+(\Delta r)^2+4 r r')^{k/2}\over (2m r r')^k}
         ~K_k(m\sqrt{ (\Delta t)^2+(\Delta r)^2+4 r r'})
         \right]  . \n{3.17}
\eea

Until now we were working with the series representation for the
four-dimensional Green function $G^0$. On the other hand, one can 
start with the two-dimensional theory defined by (\ref{3.9}). 
In order to calculate the two-dimensional quantity 
$\langle \hat{\Phi}^2(X)\rangle^{{\cal F}_{\ell}}_{\ind{ren}}$ for 
the two-dimensional operator ${\cal F}_{\ell}$, one subtracts from the  
Green function ${\cal G}_{\ell}(x,x')$ the free-field Green
function\footnote{
In two dimensions, renormalization consists of subtracting only the 
first term of the Schwinger-DeWitt expansion of the heat kernel 
for the operator ${\cal F}_{\ell}$.  In flat spacetime this is equivalent to
the subtraction of the free-field Green function, even in the presence of 
the dilaton field and a nontrivial potential $V$.
}\label{twodim}, and takes the coincidence limit:
\be\n{3.18}
\langle \hat{\Phi}^2(x)\rangle^{{\cal F}_{\ell}}_{\ind{ren}}
  =  \lim_{x'\rightarrow x} \left[ {\cal G}_{\ell}(x,x')-
         {\cal G}^{\ind{div}}_{\ell}(x,x') \right] \, ,
\ee
where the two-dimensional free-field Green function is  
\be\n{3.19}
{\cal G}^{\ind{div}}_{\ell}(x,x') 
  =  {1\over 2\pi} K_0(m\sqrt{(\Delta t)^2+(\Delta r)^2})\, .
\ee
Note that ${\cal G}^{\ind{div}}_{\ell}$ is identical to the first term in 
the first sum for ${\cal G}^0_{\ell}$ in (\ref{3.17}).

By comparing (\ref{3.11})--(\ref{3.12}) with (\ref{3.18}) we get
\be\n{3.20}
\langle \hat{\Phi}^2(x)\rangle^{{F}}_{\ind{ren}}
  =  {1\over 4\pi r^2} \sum_{\ell=0}^{\infty} (2\ell +1) \left[ \,
         \langle \hat{\Phi}^2(x)\rangle^{{\cal F}_{\ell}}_{\ind{ren}} 
         +\Delta\langle \hat{\Phi}^2(r)\rangle_{\ell} 
     \, \right] \, ,
\ee
where the anomalous term $\Delta\langle \hat{\Phi}^2\rangle_{\ell}$ is 
given by 
\bea
\Delta\langle \hat{\Phi}^2(r)\rangle_{\ell} 
  & = &  \lim_{x'\rightarrow x} \left[ \,
             {\cal G}^{\ind{div}}_{\ell}(x,x')
             -{\cal G}^0_{\ell}(x,x') \,
         \, \right]  \n{3.21a} \\
  & = &  \frac{1}{4\pi} 
             \sum_{k=1}^{\ell} {(\ell+k)!\over (\ell-k)!}
             \frac{1}{k} \frac{(-1)^{k+1}}{(mr)^{2k}} 
         +\frac{(-1)^{\ell}}{2\pi} 
             \sum_{k=0}^{\ell}{(\ell+k)!\over k!(\ell-k)!}
         ~\frac{K_k(2 m r)}{(m r)^k} \, . \n{3.21}
\eea
For example,  
\bea 
\Delta\langle \hat{\Phi}^2(r)\rangle_{\ell=0}
  & = &  \frac{1}{2\pi} K_0(2mr)  \nonumber \\ 
\Delta\langle \hat{\Phi}^2(r)\rangle_{\ell=1}
  & = &  \frac{1}{2\pi} \left[ \frac{1}{(mr)^2} - K_0(2mr) 
             -\frac{2}{(mr)}K_1(2mr) \right] \\
\Delta\langle \hat{\Phi}^2(r)\rangle_{\ell=2}
  & = &  \frac{1}{2\pi} \left[ \frac{3}{(mr)^2} - \frac{6}{(mr)^4}
             + K_0(2mr) + \frac{6}{(mr)}K_1(2mr) + \frac{12}{(mr)^2}K_2(2mr) 
         \right] \, .  \nonumber 
\eea

Relation (\ref{3.20}) explicitly demonstrates that 
$\langle \hat{\Phi}^2(x)\rangle^{{F}}_{\ind{ren}}$ can be 
expressed as a sum over modes of 
$\langle \hat{\Phi}^2(x)\rangle^{{\cal F}_{\ell}}_{\ind{ren}}$ from 
the corresponding two-dimensional theory.  However, to obtain the correct 
result we see that each term in the decomposition must be  
modified by adding the state-independent quantity 
$\Delta\langle \hat{\Phi}^2(r)\rangle_{\ell}$ of (\ref{3.21a}). 
Failure to account for these extras terms would result, for example, 
in a nonzero value for 
$\langle \hat{\Phi}^2(x)\rangle^{{F}}_{\ind{ren}}$ in the Minkowski state.

The violation of the expected mode decomposition for a physical 
observable in the process of dimensional reduction due to the 
renormalization procedure is called the dimensional-reduction anomaly.  
In this particular example the dimensional-reduction anomaly is given  
explicitly by (\ref{3.21}). There are several reasons why this anomaly arises.  
First, the number of divergent terms which are to be subtracted in the 
renormalization procedure is different in 4 and 2 dimensions (see the footnotes
on pages~\pageref{fourdim} and~\pageref{twodim}).  
However, it is easily shown that even if we make an additional 
subtraction of the next term of the Schwinger-DeWitt expansion in 
two dimensions, for $\ell\ge 2$ there remains a non-vanishing 
``local'' part of the dimensional-reduction anomaly which is given 
by the first sum in (\ref{3.21}), with $k\ge2$.  Besides this
local contribution, in our particular case there is a ``non-local'' part
to the anomaly given by the second sum.  It is present because the 
two-dimensional $(t,r)$ space is a half-plane and the two-dimensional field
vanishes at the boundary $r=0$.  While the mode-decomposed subtraction term 
${\cal G}^0_{\ell}$ from four dimensions obeys this boundary condition, the 
two-dimensional subtraction term ${\cal G}^{\ind{div}}_{\ell}$ does not.

\subsection{Rindler space}
\label{s3B}

As a second example of the dimensional-reduction anomaly in flat spacetime we 
consider the decomposition of a scalar field into Rindler time modes.
The metric for (Euclidean) Rindler space is given by
\be\n{3.22}
ds^2 = z^2 dt^2 + dz^2 + dx^2 + dy^2   ~,
\ee
where $t \in [0,2\pi)$, $z \in [0,\infty)$, $x,y \in (-\infty,\infty)$.
This line element may be obtained from the flat-space line element
\be\n{3.23}
ds^2 = dT^2 + dX^2 + dY^2 + dZ^2
\ee
by the coordinate transformation
\bea
 T = z \sin(t) \, ,  &   X = x   \, , \nonumber  \\
 Z = z \cos(t) \, ,  &   Y = y   \, , \n{3.24}
\eea
and is clearly just flat space in polar coordinates.  Hence, the 
free-field Green function is given by (\ref{3.2}) with 
\be\n{3.25}
2 \sigma = (\Delta x)^2 + (\Delta y)^2 + (\Delta z)^2 
           + 2 z z' \left[ 1 - \cos(\Delta t) \right] 
\ee
This Green function may be decomposed in terms of the Rindler ``time'' modes 
$\cos( k \Delta t)$ as an ordinary Fourier cosine series, 
\bea
G^0(X,X')  
 & = &  \frac{1}{2\pi} \frac{{\cal G}^0_{0}(\vec{x},\vec{x}')}{\sqrt{z z'}}
        +\sum_{k=1}^{\infty} \frac{\cos( k \Delta t)}{\pi}
            \frac{{\cal G}^0_{k}(\vec{x},\vec{x}')}{\sqrt{z z'}}  
        \, , \n{3.26}  \\
\frac{{\cal G}^0_k(\vec{x},\vec{x}')}{\sqrt{z z'}}
 & = &  \int_0^{2\pi}  \! \! \! d (\Delta t) \cos( k \Delta t) G^0(X,X') 
        \quad \quad ; ~ k \ge 0 \, , \n{3.27}
\eea
where $\vec{x}=(x,y,z)$. The three-dimensional Green 
function ${\cal G}^0_{k}(\vec{x},\vec{x}')$ is a solution of the equation
\be\n{3.28}
{\cal F}_{k}\, \, {\cal G}^0_k(\vec{x},\vec{x}') 
 \equiv ({}^3\Box-m^2-{V}_{k}){\cal G}^0_{k}(\vec{x},\vec{x}')
    =   -\delta(x-x') \, \delta(y-y')\, \delta(z-z') \, ,
\ee
where ${}^3\Box =\partial_x^2 +\partial_y^2 +\partial_z^2$, and
\be\n{3.29}
{V}_{k}(\vec{z})  =  \frac{4k^2-1}{4z^2} \, .
\ee
As before, we may obtain an explicit expression 
for ${\cal G}^0_{k}(\vec{x},\vec{x}')$ 
by decomposing the four-dimensional Green function, as in (\ref{3.27}).  
Using the integral representation for $K_1$ 
from Appendix~\ref{K_formulae}, combined with the integrals 
\be\n{3.31}
\int_0^{2\pi} dx \cos(kx) \exp{ \left(p\cos(x)\right) }   =  2\pi I_k(|p|)
\ee
and 
\be\n{3.32}
\int_0^{\infty} dx \frac{1}{x} \exp{ (-a x - \frac{b}{x}) } I_k(c x)
  =  2 I_k(\sqrt{b(a+c)}-\sqrt{b(a-c)}) K_k(\sqrt{b(a+c)}+\sqrt{b(a-c)}) 
\ee
(see, for example, 3.937.2 of \cite{GrRy:94} and 
2.15.6.4 of \cite{PrBrMa:86} respectively), one can show 
that  
\bea
\frac{{\cal G}^0_{k}(\vec{x},\vec{x}')}{\sqrt{z z'}} 
  & = &  -\frac{m^2}{4\pi} \left[
             \left(\vphantom{\sum}\right.
                 I_{k+1}(\alpha_-)+\frac{k}{\alpha_-}I_k(\alpha_-)
             \left.\vphantom{\sum}\right)
             K_k(\alpha_+) 
             \left(\vphantom{\sum}\right.
                 \frac{1}{md_+}-\frac{1}{md_-}
             \left.\vphantom{\sum}\right)
         \right. \nonumber \\
  &   &  \hspace{0.4in} \left. \mbox{}  
             -I_k(\alpha_-)
             \left(\vphantom{\sum}\right.
                 K_{k+1}(\alpha_+)-\frac{k}{\alpha_+}K_k(\alpha_+)
             \left.\vphantom{\sum}\right)
             \left(\vphantom{\sum}\right.
                 \frac{1}{md_+}+\frac{1}{md_-}
             \left.\vphantom{\sum}\right)  
         \right]  ,   \n{3.33}
\eea
where we have defined 
\be\n{3.34}
\alpha_\pm = \frac{m}{2}(d_+ \pm d_-) ~,
\qquad
d_\pm = \sqrt{(x-x')^2+(y-y')^2+(z\pm z')^2} ~ .
\ee

The physical observable of interest, 
$\langle \hat{\Phi}^2\rangle^F_{\ind{ren}}$, can be calculated from
the Green function $G(X,X')$ in the presence of a $t$-independent boundary
or potential using (\ref{3.3}, \ref{3.4}).  
Decomposing $G$ in the same manner as $G^0$ then allows us to write a 
decomposed form for $\langle \hat{\Phi}^2\rangle^F_{\ind{ren}}$ in 
analogy to (\ref{3.11}, \ref{3.12}):
\be\n{3.35}
\langle \hat{\Phi}^2(\vec{x})\rangle^F_{\ind{ren}}
   =   \frac{1}{\pi z} \lim_{\vec{x}'\rightarrow \vec{x}} \left\{ 
           \frac{1}{2} [ {\cal G}_{0}(\vec{x},\vec{x}')
               -{\cal G}^0_{0}(\vec{x},\vec{x}') ]
           +\sum_{k=1}^{\infty} [ {\cal G}_{k}(\vec{x},\vec{x}')
               -{\cal G}^0_{k}(\vec{x},\vec{x}') ]  \right\}  \, .
\ee
On the other hand, the renormalized value of 
$\langle \hat{\Phi}^2\rangle$ for the three-dimensional operator
${\cal F}_k$ in (\ref{3.28}) is obtained by subtracting from the full
three-dimensional Green function ${\cal G}_{k}(\vec{x},\vec{x}')$ 
not ${\cal G}^0_{k}(\vec{x},\vec{x}')$, but rather the first term in the
Schwinger-DeWitt expansion for ${\cal F}_k$, denoted 
${\cal G}^{\ind{div}}_k(\vec{x},\vec{x}')$.  Specifically, for each $k$ we have
\be\n{3.36}
\langle \hat{\Phi}^2(\vec{x})\rangle^{{\cal F}_k}_{\ind{ren}}
  =  \lim_{\vec{x}'\rightarrow \vec{x}} \left[ {\cal G}_k(\vec{x},\vec{x}')
         -{\cal G}^{\ind{div}}_k(\vec{x},\vec{x}') \right] \, ,
\ee      
where
\be\n{3.37}
{\cal G}^{\ind{div}}_k(\vec{x},\vec{x}') 
  =  \left( \frac{m}{4\pi^2 \sqrt{2\sigma}} \right)^{\frac{1}{2}}
     K_{\frac{1}{2}}(m\sqrt{2\sigma})  ~.
\ee   
Comparing (\ref{3.35}), (\ref{3.36}), we find 
\be\n{3.38}
\langle \hat{\Phi}^2(\vec{x})\rangle^F_{\ind{ren}}
   =   \frac{1}{\pi z} \left\{ \frac{1}{2} [ 
           \langle \hat{\Phi}^2(\vec{x})\rangle^{{\cal F}_0}_{\ind{ren}}
           +\Delta \langle \hat{\Phi}^2(z) \rangle_0 ]
       +\sum_{k=1}^{\infty} [ 
       \langle \hat{\Phi}^2(\vec{x})\rangle^{{\cal F}_k}_{\ind{ren}}
           +\Delta \langle \hat{\Phi}^2(z) \rangle_k ]  \right\} \, ,
\ee
where for each $k$ the anomaly is 
\bea
\Delta \langle \hat{\Phi}^2(z) \rangle_k
  & = &  \lim_{\vec{x}'\rightarrow \vec{x}} 
         \left[ {\cal G}^{\ind{div}}_k(\vec{x},\vec{x}') 
             -{\cal G}^0_{k}(\vec{x},\vec{x}')  \right]  \n{3.39}    \\
  & = &  \frac{m}{4\pi} \left[ -1 
             + m z \left( I_{k+1}(mz)K_{k+1}(mz) + I_{k}(mz)K_{k}(mz) 
             \vphantom{ \sum }    \right)
         \right.  \nonumber \\
  &   &  \left. \hspace{0.3in} \mbox{}   
         + k \left( I_{k}(mz)K_{k+1}(mz) - I_{k+1}(mz)K_{k}(mz) 
             \vphantom{ \sum }  \right)
         \right] ~. \n{3.40}  
\eea
Once again, the anomaly is seen as the difference in the subtraction terms used
to renormalize the higher- and lower-dimensional theories.  
It is easily shown that (\ref{3.40}) falls off as $O(z^{-1})$ for 
large $mz$, and diverges as $O(z^{-1})$ when $mz\rightarrow0$ for $k\ne0$.  
For $k=0$ the anomaly is finite as $mz\rightarrow0$.

These results are qualitatively very similar to those from the spherical
decomposition of flat spacetime in the previous section.  This should not be
surprising, considering that the Euclidean Rindler space (\ref{3.22}) 
is simply flat space in polar coordinates.  
As a result, the calculations of this section amount to a
one-dimensional ``spherical'' decomposition of flat spacetime.


\section{The Dimensional-Reduction Anomaly 
for $\mathbf{ \langle \hat{\Phi}^2 \rangle^{\mbox{\small{ren}}} }$}
\label{s4}
\setcounter{equation}0

In the previous section we examined the dimensional-reduction anomaly for mode
decompositions in flat space.  In the next two sections we extend these 
calculations to more general, curved spaces.  While our chief aim is the 
study of the dimensional-reduction anomaly in the effective action, 
in order to illustrate more clearly the effect of curvature on local 
reduction anomalies we begin with the simpler case of the anomaly for 
$\langle \hat{\Phi}^2\rangle^{\ind{ren}}$.  

At this point some conventions on notation are in order.  Henceforth the 
scalar field operator in four dimensions is denoted by $F$, and its Green
function is $G^F$.  The calligraphic symbols ${\cal F}_\omega$ and 
${\cal G}^{{\cal F}_\omega}$ represent the corresponding
quantities in the dimensionally-reduced theory.  In both cases the subtraction
terms for renormalization are identified by the subscript $()_{\ind{div}}$
(rather than $()^0$ as in the previous section).  Also, the mode decomposition
of $G^F_{\ind{div}}$ will be denoted by $G^F_{\ind{div}}(\omega)$ to
distinguish it more clearly from ${\cal G}^{{\cal F}_\omega}_{\ind{div}}$.

\subsection{(1+3)-reduction}
\label{s4A}

We begin with the case $n=1$ and write the metric (\ref{2.1}) in the 
form $(a,b=1,2,3)$
\be\label{4.1}
ds^2 = g_{\mu\nu}\, dX^{\mu}\, dX^{\nu} 
     = {\rm e}^{-4\phi(x)}dt^2 + h_{ab}(x)dx^a dx^b  ~.
\ee
We assume that $t \in (-\infty,\infty)$, corresponding to a 
zero-temperature state.  The scalar field operator is taken 
to be $F = \Box-V-m^2$, where the potential $V$ is independent of  Euclidean 
``time'' $t$.  It is easy to see that the
operator $\Delta_{\Omega}$ for the metric (\ref{4.1}) is
$\partial^2/\partial t^2$. Hence, the mode decomposition in terms of its
eigenvalues is simply the standard Fourier transform with
\be\label{4.2}
Y(t) = \exp(\pm i \omega t) \, .
\ee

The bare $\langle \hat{\Phi}^2\rangle$ is obtained from the coincidence limit 
of the Green function, 
\be
\langle \hat{\Phi}^2 (X) \rangle^F
  =  \lim_{ X' \rightarrow X } G^F(X,X') \, . \label{4.3}
\ee
For a general four-dimensional space, 
the divergences of the Green function in the
coincidence limit come from the first two terms of the
Schwinger-DeWitt expansion of the heat kernel,  
\be\label{4.4}
G^F_{\ind{div}}(t,x;t',x') 
    =   \int_0^\infty ds \frac{1}{(4\pi s)^2} 
         \exp{ \left\{ - m^2s - \frac{2\sigma + \epsilon^2}{4s} \right\} }  
         \left[ \Re^{\Box-V}_0 + s\Re^{\Box-V}_1 \right]  ~.
\ee
Here $\sigma = \sigma(X,X')$ is one-half of the square of the
geodesic distance between points $X=(t,x^a)$ and $X'=(t', x'\mbox{}^a)$, and
\be \label{4.5}
\Re^{\Box-V}_n = \Delta^{\frac{1}{2}}(X,X')\, 
                 \, a_n^{\Box-V}(X,X') \, ,
\ee
where $\Delta(X,X')$ is the Van Vleck determinant, 
\be\label{vanvleck}
\Delta(X,X') = \frac{1}{\sqrt{g(X)}\sqrt{g(X')}}
               \det \left[ \frac{\partial\hphantom{X^\mu}}{\partial X^\mu}
                   \frac{\partial\hphantom{X^\nu}}{\partial X^\nu} 
                   \sigma(X,X') \right] \, ,
\ee
and the first few Schwinger-DeWitt coefficients  
in the coincidence limit $X' \rightarrow X$ are 
\bea
a_0^{\Box-V}  & = &  1  \, , \nonumber \\
a_1^{\Box-V}  & = &  \frac{1}{6}\,^4\!R - V  \, ,  \\
a_2^{\Box-V}  & = &  \frac{1}{180} \left[ \,^4\!R_{\alpha\beta\gamma\delta}
                         \,^4\!R^{\alpha\beta\gamma\delta}
                         -\,^4\!R_{\alpha\beta} \,^4\!R^{\alpha\beta} 
                     \right]  
                     +\frac{1}{2} \left( \vphantom{\sum} \right.
                         \frac{1}{6} \,^4\!R - V 
                     \left. \vphantom{\sum} \right)^2  
                     +\frac{1}{30} \Box\,^4\!R 
                     -\frac{1}{6} \Box V  \, . \nonumber  
\eea
Expansions of $\sigma$ and the $\Re^{\Box-V}_n$ for the metric (\ref{4.1}) 
with $x=x'$ and $\e^{-2\phi}(t-t')$ small are given in 
Appendix~\ref{1+3_expansions}.

Clearly, the Green function and Schwinger-DeWitt coefficients depend on
the form of the field equation.  We make this dependence explicit by
providing such quantities with a corresponding superscript.  The superscript 
for the $a_n$ does not include the mass term from the operator $F$, as it is
accounted for separately in (\ref{4.4}).
We also include a cut-off parameter $\epsilon$ in the exponent of the
integrand in (\ref{4.4}) which ensures convergence of the integral for
small $s$ with $\sigma$ vanishing.

Our purpose is to compare the divergences of four- and 
three-dimensional theories related by a Fourier time transform.
Unfortunately, it is not possible to evaluate the Fourier transform of
(\ref{4.4}) exactly for general $h_{ab}$.  Our response is to 
make point splitting in the $t$-direction, expanding all 
$t$-dependent quantities in powers of the curvature, and truncating all
expressions at first order in the curvature 
(two derivatives of the dilaton or metric).  
Denoting $\tau \equiv \e^{-2\phi} (t-t')$
and putting $x=x'$ we have up to first order in the curvature  
\bea
2\sigma(t,x;t',x)  
  & = &  \tau^2 - \frac{1}{3}(\nabla\phi)^2 \tau^4 \, , \label{4.6}   \\
\Re^{\Box-V}_0
  & = &  1 + \frac{1}{6}\Box\phi \, \tau^2  \, ,   \label{4.7}  \\
\Re^{\Box-V}_1
  & = &  \frac{1}{6}R - V + \frac{2}{3}\Box\phi  ~. \label{4.8}  
\eea
See Appendix \ref{1+3_expansions} for details.
Here $R$ is understood to be the scalar curvature for the
three-metric $h$, and is related to the four-dimensional 
curvature $ ^4\!R$ via  
\be \label{4.10}
^4\!R  =  R + 4\Box\phi \, .
\ee  
Substituting (\ref{4.6}) into (\ref{4.4}), expanding the exponent and
keeping in the exponent only terms which are quadratic in $\tau$, we get
\be \label{4.11}
G^F_{\ind{div}}(t,x;t',x) 
  =   \int_0^\infty  \frac{ds}{(4\pi s)^2} 
         \exp{ \left\{ - m^2s - \frac{\tau^2 + \epsilon^2}{4s} \right\} }
         \left[ \left( 1 + \frac{(\nabla\phi)^2}{12s}\tau^4 \right)
         \Re^{\Box-V}_0 + s\Re^{\Box-V}_1 \right] \, . 
\ee
The integral over the parameter $s$ can be taken with the following
result (see Appendix~\ref{K_formulae}):
\bea
G^F_{\ind{div}}(t,x;t',x)
  & = &  \frac{1}{8\pi^2} \left[ \;
         \left( \frac{1}{6}R-V+\frac{2}{3}\Box\phi \right) {\bf K}_0(z)
         +m^2 \left(1 + \frac{\Box\phi}{6} \tau^2 \right) {\bf K}_1(z)
         \right. \nonumber \\
  &   &  \hspace{0.4in} \mbox{} 
         +m^4 \frac{(\nabla\phi)^2}{12}\tau^4 {\bf K}_2(z)
         \; \left. \vphantom{\frac{1}{1}} \right] \, . \label{4.12}
\eea  
Here $z \equiv m\sqrt{\tau^2+\epsilon^2}$, 
\be\n{4.12a}
{\bf K}_{\nu}(z)=\left( {2\over z}\right)^{\nu} \, K_{\nu}(z)\, ,
\ee
and the $K_\nu$ are modified Bessel functions.  Putting $\epsilon=0$ 
and expanding $G^F_{\ind{div}}$ in a Laurent series in $\tau$ we get
\bea
G^F_{\ind{div}}(t,x;t',x)
  & = &  {1\over 4\pi^2\tau^2} 
         +{1\over 8\pi^2}\left[ m^2 
             -\left({1\over 6}R-V+{2\over 3}\Box\phi\right)\right]\, 
             \left\{ \gamma + {1\over 2}\ln \left({m^2\tau^2\over 4}\right)
                 \right\}
         \nonumber \\
  &   &  \mbox{}
         -{m^2\over 16\pi^2} + {(\nabla\phi)^2\over 12\pi^2} 
         +{\Box\phi \over 24\pi^2} + \ldots \, . \label{4.13}
\eea
Here $\gamma$ is the Euler constant, and the dots denote terms of higher
order in $\tau$.  The terms displayed are just the usual DeWitt-Schwinger 
expansion for the divergent parts of $G^F$ for the metric (\ref{4.1}).

The renormalized value of $\langle \hat{\Phi}^2\rangle^F$ can be 
written in the form
\be\label{4.14}
\langle \hat{\Phi}^2(t,x)\rangle_{\ind{ren}}^F 
  =  \lim_{\epsilon\rightarrow 0} \lim_{t-t'\rightarrow 0}
     \left[ G^F(t,x;t',x) - G^F_{\ind{div}}(t,x;t',x) \right]\, .
\ee
Taking the limit $\epsilon \rightarrow 0$ in this expression is a trivial
operation since the difference in the square brackets is already a
finite quantity.

Let us now analyze what happens when we mode decompose $G^F_{\ind{div}}$ 
and compare to the corresponding divergent terms from the three-dimensional 
theory.  The Fourier time-transform pair is defined as 
\bea
G^F_{\ind{div}}(x;x'|\omega) 
  & = &  \e^{-(\phi+\phi')} \int_{-\infty}^\infty d(t-t') \e^{i\omega (t-t')} 
         G^F_{\ind{div}}(t,x;t',x')  \, , \label{4.15}  \\
G^F_{\ind{div}}(t,x;t',x')
  & = &  \e^{(\phi+\phi')} \int_{-\infty}^\infty \frac{d\omega}{2\pi}
         \e^{-i\omega (t-t')} G^F_{\ind{div}}(x;x'|\omega) \, . \label{4.15b}
\eea
Since $G^F_{\ind{div}}(t,x;t',x')$ depends only on the difference
$t-t'$, the function $G^F_{\ind{div}}(x;x'|\omega)$ does not depend
on $t$ and $t'$. 
Calculating the integral in (\ref{4.15}) using (\ref{4.12}) and (\ref{3int5}) 
we obtain
\be\label{4.16}
G^F_{\ind{div}}(x;x|\omega)  
   =   \frac{1}{4\pi}  
       \left[ \;
           \frac{1}{\epsilon} - \mu 
           + \frac{1}{2\mu} \left( \frac{1}{6}R-V+\frac{2}{3}\Box\phi \right) 
           + \frac{m^2\Box\phi}{6\mu^3} 
           + \frac{m^4(\nabla\phi)^2}{2\mu^5} 
           + O(\epsilon)  \; \right]   \, ,
\ee
where $\mu \equiv \sqrt{m^2+\e^{4\phi}\omega^2}$.  Meanwhile, 
the operator ${\cal F}_{\omega}$ which determines the reduced
equation of motion (\ref{2.9}) is 
\be\label{4.17}
{\cal F}_{\omega} = \Delta_h-V_{\omega}[\phi]-m^2\, ,
\ee
where
\be\label{4.18}
V_{\omega}[\phi] 
  =  \omega^2 {\rm e}^{4\phi} + (\nabla\phi)^2 - \Delta_h\phi  + V \, .
\ee
The divergent part of the Green function for the
operator ${\cal F}_{\omega}$ in three dimensions is 
generated by the first term in the
Schwinger-DeWitt expansion of the heat kernel and is 
\be\label{4.19}
{\cal G}^{{\cal F}_{\omega}}_{\ind{div}}(x;x)
  =  \int_0^\infty ds 
         \frac{1}{(4\pi s)^{\frac{3}{2}}} 
         \exp{ \left\{ - m^2s - \frac{\epsilon^2}{4s} \right\} }
  =  \frac{1}{4\pi} \left[ \; \frac{1}{\epsilon} - m + O(\epsilon) 
         \; \right] \, .
\ee
Hence, if we start with a three-dimensional theory with the field
equation
\be\label{4.20}
{\cal F}_{\omega} \hat{\Phi}(x) = 0\, ,
\ee
we will obtain for the renormalized 
value of $\langle \hat{\Phi}^2\rangle^{{\cal F}_{\omega}}$ the 
representation
\be\label{4.21}
\langle\hat{\Phi}^2(x)\rangle_{\ind{ren}}^{{\cal F}_{\omega}}
  =  \lim_{\epsilon\rightarrow 0}\left[
         {\cal G}^{{\cal F}_{\omega}}(x;x)
         -{\cal G}^{{\cal F}_{\omega}}_{\ind{div}}(x;x) 
     \right]\, .
\ee
By comparing (\ref{4.14}) with (\ref{4.21}) we can get the following
relation between $\langle \hat{\Phi}^2 \rangle$ in the 
four- and three-dimensional theories:
\be \label{4.22}
\langle \hat{\Phi}^2 \rangle^F_{\ind{ren}} 
  =  \e^{2\phi} \int_{-\infty}^\infty \frac{d\omega}{2\pi} \left[ 
         \langle \hat{\Phi}^2 \rangle^{{\cal F}_{\omega}}_{\ind{ren}} 
         +\Delta \langle \hat{\Phi}^2 \rangle_{\omega} 
     \right]\, ;
\ee
where 
\bea \label{4.23}
\Delta \langle \hat{\Phi}^2 \rangle_{\omega}  
  & = &  \lim_{\epsilon\rightarrow0} \left[ 
             {\cal G}^{{\cal F}_{\omega}}_{\ind{div}}(x;x)
             -G^F_{\ind{div}}(x;x|\omega)
         \right]   \\
  & = &  \frac{1}{4\pi} \left[ \;
             \mu - m  
             -\frac{1}{2\mu} \left( \frac{1}{6}R-V+\frac{2}{3}\Box\phi \right) 
             -\frac{m^2 \, \Box\phi}{6\mu^3} - \frac{m^4(\nabla\phi)^2}{2\mu^5} 
         \; \right]  
\eea
(compare to (\ref{2.23})).
Since $\Delta \langle \hat{\Phi}^2 \rangle_{\omega}$ does not vanish we
have another example of the dimensional-reduction anomaly.

The anomalous term $\Delta \langle \hat{\Phi}^2\rangle_{\omega}$ is finite
and can be written in the form
\be\label{4.24}
\Delta \langle \hat{\Phi}^2\rangle_{\omega} 
  =  \Delta \langle \hat{\Phi}^2\rangle_{\omega}^\sharp   
     +\Delta \langle \hat{\Phi}^2\rangle_{\omega}^\flat \, ,  
\ee
where
\be\label{4.25}
\Delta \langle \hat{\Phi}^2\rangle_{\omega}^\sharp 
  =  \frac{1}{4\pi} \left[ \; 
         \bar{\omega} + \frac{m^2}{2\bar{\omega}} - m 
         -\frac{1}{2\bar{\omega}} \left( \frac{1}{6}R-V+\frac{2}{3}\Box\phi
              \right) 
     \; \right]  \, ,
\ee
\be\label{4.26}
\Delta\langle \hat{\Phi}^2\rangle_{\omega}^\flat 
  =  \frac{1}{4\pi} \left[ \; 
         \mu - \bar{\omega} - \frac{m^2}{2\bar{\omega}} 
         - \frac{1}{2} \left( \frac{1}{6}R-V+\frac{2}{3}\Box\phi \right)
             \left(\frac{1}{\mu}-\frac{1}{\bar{\omega}}\right) 
         - \frac{m^2 \, \Box\phi}{6\mu^3} 
         - \frac{m^4(\nabla\phi)^2}{2\mu^5} \; \right] \, .
\ee
Here and later we use the notation $\bar{\omega}=\e^{2\phi}\omega$.  
The quantity $\Delta \langle \hat{\Phi}^2\rangle_{\omega}^\sharp$ is
that part of the anomaly which dominates at high frequency $\omega$; it
consists of all terms of $O(\omega^{-1})$ and higher in the
large-$\omega$ expansion of $\Delta \langle \hat{\Phi}^2\rangle_{\omega}$. 
These are the terms which diverge in the integration over $\omega$, 
and hence which lead to the divergences in the four-dimensional 
Green function as $t-t'\rightarrow 0$. The part of the anomaly that 
remains when $\Delta \langle \hat{\Phi}^2\rangle_{\omega}^\sharp$ is 
subtracted off is denoted by 
$\Delta\langle \hat{\Phi}^2\rangle_{\omega}^\flat$ 
\footnote{The symbol $\sharp$ ({\it sharp}) is borrowed 
from musical notation to denote the {\it high-frequency} part of the
anomaly; the remainder is labelled with $\flat$ ({\it flat}).}. 
It is of $O(\omega^{-1})$ for high frequencies, so the Fourier 
transform of $\Delta\langle \hat{\Phi}^2\rangle_{\omega}^\flat$  
is finite as $t-t'\rightarrow 0$.

It should be emphasized that, generally speaking, the procedure of
determining the Fourier transform for the point-split divergent
part of the Green function is not unique.  The short distance
Schwinger-DeWitt expansion guarantees us only the correct reproduction
of terms which are singular in the coincidence limit $t-t'\rightarrow 0$.
For this reason, only the high-frequency 
part $\langle \hat{\Phi}^2\rangle_{\omega}^{\sharp}$
which is responsible for the short distance behavior of
$G^F_{\ind{div}}(t,x;t',x) $ in four dimensions is a universal
function. A particular form of the low frequency part of the
anomalous term,  $\langle \hat{\Phi}^2\rangle_{\omega}^{\flat}$, may depend on
the concrete form of the continuation of $G^F_{\ind{div}}(t,x;t',x) $
for large separation of points $t$ and $t'$. Our choice (\ref{4.12}), though
not unique, possesses a number of pleasant 
properties. First of all, in a flat spacetime it is identical to the
Green function of the scalar field for both the massive and massless
cases.
Moreover, the  reduction anomaly obtained  by using this prescription is
directly related with a so-called {\em analytic approximation} for
$\langle \hat{\Phi}^2\rangle^F_{\ind{ren}}$ derived in \cite{Ande:90}.
In order to demonstrate this let us take the inverse Fourier transform of
$\langle \hat{\Phi}^2\rangle_{\omega}^{\flat}$,
\be\label{4.27}
\langle \hat{\Phi}^2 \rangle^F_{\ind{approx}} 
  =  \int_0^\infty \frac{d\bar{\omega}}{\pi} 
     \Delta \langle \hat{\Phi}^2\rangle_{\omega}^{\flat} \, .
\ee
Performing the integration (see Appendix \ref{mu_integrals}), we obtain
\be\label{4.28}
\langle \hat{\Phi}^2 \rangle^F_{\ind{approx}} =
\frac{m^2}{16\pi^2} 
     +\frac{1}{16\pi^2} 
         \left( \frac{1}{6}\,^4\!R-V -m^2 \right) 
         \ln{ \frac{m^2 \e^{-4\phi}}{4\eta^2} }
         - \frac{\Box\phi}{24\pi^2} - \frac{(\nabla\phi)^2}{12\pi^2} ~.  
\ee 
The parameter $\eta$ is a low-frequency cut-off which is required to make
the integral convergent. It corresponds to a well-known ambiguity in the
renormalization prescription. This ambiguity is absent for a 
conformally-invariant theory, when $m=0$ and the parameter of the
non-minimal coupling $\xi$ takes its conformal value $\xi=1/6$.

For the special case of a static spherically symmetric spacetime this
reproduces exactly the analytic approximation of Anderson
\cite{Ande:90}.  Relation (\ref{4.28}) also reproduces the
zero-temperature  Killing approximation \cite{FrZe:87} for a massless
conformally-coupled field in a static spacetime.  Relation (\ref{4.28})
can be considered in fact as an extension of these results to the
general case when the spacetime is static, but not necessary spherically
symmetric, and the field equation includes an arbitrary mass and
potential $V$.  We shall discuss the origin of this approximation,
its finite temperature generalization,  and its properties elsewhere 
\cite{FrSuZe:99}.

\subsection{(2+2)-reduction}
\label{s4B}

Let us discuss now the dimensional-reduction anomaly for $\langle
\hat{\Phi}^2\rangle^{\ind{ren}}$ for the case when the metric of the 
internal space is flat and two-dimensional; that is, the spacetime 
metric is of the form $(A,B=2,3)$
\be\label{4.29}
ds^2 = {\rm e}^{-2\phi(x)}(dt_0^2 +dt_1^2)+ h_{AB}(x)dx^A dx^B  \, .
\ee
First of all, it should be noticed that this metric is a special case
of the metric (\ref{4.1}) when the dilaton field does not depend on one
of the coordinates $x^a$.  Equation (\ref{4.29}) can be obtained 
from (\ref{4.1}) by rescaling the dilaton field $\phi \rightarrow \phi/2$ 
and putting
\be\label{4.30}
h_{ab}dx^a dx^b ={\rm e}^{-2\phi(x)}\, dt_1^2+ h_{AB}(x)dx^A dx^B\, .
\ee
For the divergent part of the four-dimensional Green function expanded to 
first order in the curvature we have an expression similar to (\ref{4.12}),
\bea \label{4.31}
G^F_{\ind{div}}({\bf t},x;{\bf t}',x)
  & = & \frac{1}{8\pi^2} \left[ \;
             m^2 {\bf K}_1(z)
             + \frac{(\nabla\phi)^2}{48} m^4 \tau^4 {\bf K}_2(z)  
             + \frac{\Box\phi}{12} m^2 \tau^2 {\bf K}_1(z) 
             \right. \nonumber \\
  &   &      \hspace{0.4in} \left. \vphantom{\frac{(\nabla\phi)^2}{1}} \mbox{} 
             + \left( \frac{1}{6}R-V+\frac{2}{3}\Box\phi
                 +\frac{1}{3}(\nabla\phi)^2 \right) {\bf K}_0(z)
	 \right] \,  .
\eea
See Appendix~\ref{2+2_expansions}.
Here $R$ refers to the curvature of the two-dimensional space with metric 
$h_{AB}$, and we define 
\be\label{4.32}
     z  =  m\sqrt{\tau^2+\epsilon^2}\, ,\hspace{0.5cm}
\tau^2  =  \mbox{e}^{-2\phi}{\bf t}^2\equiv  \mbox{e}^{-2\phi}(t_0^2+t_1^2)\, .
\ee

To obtain the mode decomposition of the divergent part of the Green
function we make a Fourier transform similar to (\ref{4.15})
\bea
G^F_{\ind{div}}(x;x'|p) 
  & = &  \e^{-(\phi+\phi')} \int_{-\infty}^\infty d({\bf t-t'}) 
         \e^{i{\bf p}({\bf t}-{\bf t}')}
         G^F_{\ind{div}}({\bf t},x;{\bf t}',x')
         \, , \label{4.33}  \\
G^F_{\ind{div}}({\bf t},x;{\bf t}',x')
  & = &  \e^{(\phi+\phi')} \int_{-\infty}^\infty \frac{d{\bf p}}{(2\pi)^2} 
         \e^{-i{\bf p}({\bf t}-{\bf t}')}
         G^F_{\ind{div}}(x;x'|p) 
         \, . \label{4.33b}
\eea
Here we use the vector notations ${\bf p}=(p_0,p_1)$ and  
${\bf p}{\bf t}=p_0t_0 +p_1 t_1$.  We also denote $p^2={\bf p}^2$. Since the
function  $G^F_{\ind{div}}$ depends only on the difference 
${\bf t}-{\bf t}'$, its Fourier transform depends only on $p$. Calculating
the integrals we obtain
\bea
G^F_{\ind{div}}(x;x'| p)  
 & = &  \frac{1}{2\pi} \left[ \;
           -\left\{ \gamma  
           +\frac{1}{2} \ln{ \left( \frac{\mu^2\epsilon^2}{4} \right) } \right\}
           +\frac{1}{2\mu^2} \left( \frac{1}{6}R+\frac{2}{3}\Delta\phi
               -(\nabla\phi)^2-V \right)   
           \right.   \nonumber  \\
 &   &     \left. \hspace{0.35in} \mbox{}
           + \frac{m^2}{6\mu^4} \Box\phi
           + \frac{m^4}{3\mu^6} (\nabla\phi)^2 
        \; \right] + O(\epsilon) \, ,  \label{4.34} 
\eea
where
\be\label{4.35}
\mu=\sqrt{m^2+\bar{p}^2}\, ,\hspace{0.5cm}
\bar{p}=\mbox{e}^{\phi}\, p \, .
\ee

The adopted mode expansion into plane waves $\exp(i{\bf p}{\bf
t})$ reduces the initial problem for the four-dimensional
operator $F = \Box-V-m^2$ to two-dimensional problems with a
dilaton-dependent potential.  The corresponding wave operator ${\cal F}_p$ is
\be\label{4.36}
{\cal F}_{p}=\Delta_h-V_{p}[\phi]-m^2\, ,
\ee
where
\be\label{4.37}
V_{p}[\phi]  =  \bar{p}^2 + (\nabla\phi)^2 - \Delta_h\phi  + V \, .
\ee
The divergent part of the two-dimensional Green function for the
operator ${\cal F}_{p}$ can be obtained by using the Schwinger-DeWitt
expansion for this operator. In two dimensions this divergent part
is generated by only the first term in the Schwinger-DeWitt expansion 
of the heat kernel and is 
\be\label{4.38}
{\cal G}^{{\cal F}_{p}}_{\ind{div}}(x;x)
  =  \int_0^\infty ds \frac{1}{4\pi s} 
         \exp{ \left\{ -m^2s - \frac{\epsilon^2}{4s} \right\} } 
  =  -\frac{1}{2\pi} \left\{ \gamma  
         +\frac{1}{2} \ln{ \left( \frac{m^2\epsilon^2}{4} \right)  }
         \right\} + O(\epsilon)   
 \, .
\ee
We define the four- and two-dimensional renormalized 
values $\langle \hat{\Phi}^2(x)\rangle^F_{\ind{ren}}$ and 
$\langle \hat{\Phi}^2(x)\rangle^{{\cal F}_{p}}_{\ind{ren}}$ by 
expressions similar to (\ref{4.14}) and (\ref{4.21}) respectively. 
By comparing these definitions, and using relations (\ref{4.34}) 
and (\ref{4.38}) we obtain the representation
\be \label{4.39}
\langle \hat{\Phi}^2 \rangle^F_{\ind{ren}} 
    = \e^{2\phi} \int \frac{d{\bf p}}{(2\pi)^2} \left[ 
          \langle \hat{\Phi}^2 \rangle^{{\cal F}_{p}}_{\ind{ren}} 
          +\Delta \langle \hat{\Phi}^2 \rangle_{p} 
      \right]\, ,
\ee
where the dimensional-reduction anomaly 
$\Delta \langle \hat{\Phi}^2 \rangle_{p}$ is
\bea 
\Delta \langle \hat{\Phi}^2 \rangle_{p}  
& = &  \frac{1}{4\pi} \left[  \;
         \ln{ \left( \frac{\mu^2}{m^2} \right) } 
         -\frac{1}{\mu^2} \left( \frac{1}{6}R+\frac{2}{3}\Delta\phi
             -(\nabla\phi)^2-V \right)
         \right. \nonumber \\
  &   &  \left. \hspace{0.4in} \mbox{}
         -\frac{m^2}{3\mu^4} \left(\Delta\phi-2(\nabla\phi)^2 \right) 
         -\frac{2m^4}{3\mu^6} (\nabla\phi)^2
         \; \right]  \label{4.40} \\
  & = &  \frac{1}{4\pi} \left[ \;
         \ln{ \left( \frac{\mu^2}{m^2} \right) } 
         - \frac{1}{\mu^2} \left( \frac{1}{6}\,^4\!R-V \right)
         - \frac{m^2}{3\mu^4}\Box\phi 
         - \frac{2m^4}{3\mu^6} (\nabla\phi)^2
         \; \right] ~.  \label{4.41}
\eea
For convenience, we give here two different forms of the representation
for $\Delta \langle \hat{\Phi}^2 \rangle_{p} $. In the first equality,
(\ref{4.40}), all the quantities such as the curvature, covariant
derivatives, and so on are two-dimensional objects calculated for
two-metric $h_{AB}$. In the second equality, (\ref{4.41}), the same
exression is written in terms of four-dimensional objects defined for
the four-metric (\ref{4.29}).

The part of $\Delta \langle \hat{\Phi}^2 \rangle_{p} $ which dominates at 
large ``momentum'' $p$ and which is responsible for the
divergences of the four-dimensional Green function in the coincidence
limit ${\bf t}-{\bf t}'\rightarrow 0$ is
\be\label{4.42}
\Delta \langle \hat{\Phi}^2\rangle_{p}^{\sharp}
  =  \frac{1}{4\pi} \left[          
         \ln{ \frac{\bar{p}^2}{m^2} } 
         -\frac{1}{\bar{p}^2}\left(\frac{1}{6}\,^4\!R-V-m^2 \right)
     \right] \, .
\ee
Defining the sub-leading part 
$\Delta \langle \hat{\Phi}^2\rangle_{p}^{\flat}$ of the anomaly and
$\langle \hat{\Phi}^2 \rangle^F_{\ind{approx}}$
by relations similar to (\ref{4.24}) and (\ref{4.27}), respectively, we
obtain an expression for $\langle \hat{\Phi}^2 \rangle^F_{\ind{approx}}$
which is identical to (\ref{4.28}). One can expect this result, since
the $(2+2)$ reduction may be considered as a special case of the $(3+1)$
reduction.

\section{The Dimensional-Reduction Anomaly for the Effective Action}
\label{s5}
\setcounter{equation}0

\subsection{(1+3)-reduction}
\label{s5A}

For the static spacetime (\ref{4.1}), the calculation of the anomaly in
the effective action $W^F$ 
\be\label{5.1}
W^F[g]  =  \int dX \sqrt{g} \, L^F 
\ee
proceeds analogously to the calculation of the anomaly in 
$\langle \hat{\Phi}^2\rangle$. To analyse the divergent part 
of the effective action
we introduce first a point-split version of $L^F$.  Using the
Schwinger-DeWitt expansion for the heat kernel, we have for the
divergent part of $L^F$
\be\label{5.2}
L^F_{\ind{div}}(t,x;t',x)
  =  -\frac{1}{2} \int_0^\infty \frac{ds}{s} \frac{1}{(4\pi s)^2}
     \exp{ \left\{ - m^2s - \frac{2\sigma + \epsilon^2}{4s} \right\} }
     \left[ \Re^{\Box-V}_0 + s\Re^{\Box-V}_1 + s^2\Re^{\Box-V}_2 \right] ~.
\ee
As earlier, the points are split in the $t$ direction. Since the
internal space is homogeneous, the point-split Lagrangian depends on
$t-t'$. 

Faced with the same problem as before, we expand $\sigma$ 
and the $\Re^{\Box-V}_n$ in terms of $\tau \equiv \e^{-2\phi} (t-t')$ 
for $x=x'$, this time truncating at second order in the curvature 
(four derivatives of the metric or dilaton).  Writing  
\bea \label{5.3}
2\sigma(t,x;t',x)   
  & = &  \tau^2 + u \tau^4 + v \tau^6 + \cdots ~,  \nonumber \\ 
\Re^{\Box-V}_n(t,x;t',x)
  & = &  \Re^{\Box-V}_{n(0)} + \Re^{\Box-V}_{n(2)} \tau^2 
         +\Re^{\Box-V}_{n(4)} \tau^4 + \cdots   ~, 
\eea
we have $\Re^{\Box-V}_{0(0)}=1$, while $u$, $v$, and the 
other $\Re^{\Box-V}_{n(k)}$ may be found in Appendix~\ref{1+3_expansions}.  
Inserting and truncating at $O(R^2)$ gives
\bea
L^F_{\ind{div}}(t,x;t',x)  
  & = &  -\frac{1}{(4\pi)^2} \left[ \,
             m^4 {\bf K}_2(z) 
             -\frac{1}{4} (u\tau^4+v\tau^6)m^6 {\bf K}_3(z)
             +\frac{1}{32} u^2\tau^8 m^8 {\bf K}_4(z)
 	     \right. \nonumber \\
  &   &      \hspace{0.65in} \mbox{}     
             +\Re^{\Box-V}_{0(2)} \tau^2 m^4
                 \left( {\bf K}_2(z)-\frac{1}{4}u\tau^4 m^2 {\bf K}_3(z)\right) 
  	     +\Re^{\Box-V}_{0(4)} \tau^4 m^4 {\bf K}_2(z)
 	     \nonumber \\
  &   &      \hspace{0.65in} \mbox{}     
 	     +\Re^{\Box-V}_{1(0)} m^2 \left( 
	         {\bf K}_1(z)- \frac{1}{4} u\tau^4 m^2 {\bf K}_2(z) \right)
             +\Re^{\Box-V}_{1(2)}\tau^2 m^2 {\bf K}_1(z)
 	     \nonumber \\
  &   &      \hspace{0.65in} \left. \vphantom{\frac{1}{1}} \mbox{}     
	     +\Re^{\Box-V}_{2(0)} {\bf K}_0(z)  
         \, \right] \, . \label{5.4} 
\eea
The function ${\bf K}_{\nu}(z)$ is defined by (\ref{4.12a}).

We define the Fourier transform of $L^F_{\ind{div}}$ as in (\ref{4.15}).
Evaluating the transform as before yields
\bea
L^F_{\ind{div}}(x|\omega)
  & = &  -\frac{1}{4\pi}  
         \left\{  \vphantom{ \left[ \frac{1}{4\mu} \right] }
             \frac{1}{\epsilon^3}
             +\frac{1}{2\epsilon} \left[ - 3u + 2\Re^{\Box-V}_{0(2)}
                 + \Re^{\Box-V}_{1(0)} - \mu^2 \right]   
             \right. \nonumber \\
  &   &      \hspace{0.45in} \mbox{} 
             +\frac{1}{3}\mu^3 
             -\frac{1}{2} \mu \left( - 3u + 2\Re^{\Box-V}_{0(2)} 
                 +\Re^{\Box-V}_{1(0)} \right) 
             \nonumber \\
  &   &      \hspace{0.45in} \mbox{} 
             +\frac{5u\bar{\omega}^2}{2\mu}
             +\frac{m^2u\bar{\omega}^2}{2\mu^3}
             -\frac{15 m^6 v}{2\mu^7}
             +\frac{105 m^8 u^2}{8\mu^9}   
             \nonumber \\
  &   &      \hspace{0.45in} \mbox{} 
             +\Re^{\Box-V}_{0(2)} \left[ 
                 \mbox{} - \frac{\bar{\omega}^2}{\mu} 
                 -\frac{15 m^6 u}{2\mu^7} \right]   
             +\Re^{\Box-V}_{0(4)} \left[ \frac{3m^4}{\mu^5} \right] 
             \nonumber \\  
  &   &      \hspace{0.45in} \left. \mbox{} 
             +\Re^{\Box-V}_{1(0)} \left[ -\frac{3 m^4 u}{4\mu^5} \right] 
             + \Re^{\Box-V}_{1(2)} \left[ \frac{m^2}{2\mu^3}  \right] 
             + \Re^{\Box-V}_{2(0)} \left[ \frac{1}{4\mu}  \right] 
             + O(\epsilon)  
         \right\}  . \label{5.6}
\eea

Meanwhile, for the three-dimensional theory with the field operator
${\cal F}_{\omega}$ given by (\ref{4.17}) we have 
\bea
L^{{\cal F}_{\omega}}_{\ind{div}}(x)
  & = &  -\frac{1}{2} \int_0^\infty \frac{ds}{s} 
         \frac{1}{(4\pi s)^{\frac{3}{2}}}
         \exp{ \left\{ - m^2s - \frac{\epsilon^2}{4s} \right\} }
         \left[ \Re^{\Delta_h-V_{\omega}}_0 
             +s\Re^{\Delta_h-V_{\omega}}_1 \right]  
         \nonumber \\
  & = &  -\frac{1}{4\pi} \left[ \frac{1}{\epsilon^3} + 
         \frac{1}{2\epsilon} \left( \frac{1}{6}R-V_{\omega}-m^2 \right)
         + \frac{m^3}{3} - \frac{m}{2} \left( \frac{1}{6}R-V_{\omega} \right)
         \right] + O(\epsilon) \, . \label{5.7}
\eea
Here $V_{\omega}$ is the effective potential of the three-dimensional
system, and is given by (\ref{4.18}).

The renormalized effective actions in four and three dimensions
are obtained by subtracting from the exact effective action 
its divergent part, as given by (\ref{5.4}) and (\ref{5.7}) respectively. 
By comparing these divergent parts using (\ref{4.15}, \ref{4.15b}) 
and (\ref{5.6}) we can write
\be \label{5.8}
L^F_{\ind{ren}}(x) 
  =  \e^{2\phi} \int_{-\infty}^\infty \frac{d\omega}{2\pi} \left[ 
         L^{{\cal F}_{\omega}}_{\ind{ren}}(x) +\Delta L_{\omega}(x) \right]\, ,
\ee
where $\Delta L_{\omega}$ is the term representing the
dimensional-reduction anomaly of the renormalized effective action,
\bea
\Delta L_{\omega}   
  & = &  \frac{1}{4\pi} \left\{ \vphantom{ \frac{m^6}{\mu^7} } 
             \frac{1}{3}(\mu^3-m^3) 
             -\frac{1}{2} (\mu-m) \left( \frac{1}{6}R-V-
                 (\nabla\phi)^2 + \Delta_h\phi \right)        
             -\frac{1}{2} m \bar{\omega}^2  
             \right. \nonumber  \\     
  &   &      \hspace{0.3in} \mbox{} 
             +\mu \left( \frac{5}{2}u - \Re^{\Box-V}_{0(2)} \right)
             +\frac{1}{\mu} \left( -2m^2u + m^2\Re^{\Box-V}_{0(2)} 
                 +\frac{1}{4}\Re^{\Box-V}_{2(0)} \right)
             \nonumber  \\
  &   &      \hspace{0.3in} \mbox{} 
             +\frac{m^2}{\mu^3} \left( -\frac{m^2}{2}u 
                 + \frac{1}{2}\Re^{\Box-V}_{1(2)} \right)
             +\frac{m^4}{\mu^5} \left( 3\Re^{\Box-V}_{0(4)} 
                 -\frac{3}{4}u \Re^{\Box-V}_{1(0)} \right)
             \nonumber  \\
  &   &      \hspace{0.3in}  \left. \mbox{}
             +\frac{m^6}{\mu^7} \left( 
                 -\frac{15}{2} v -\frac{15}{2}u \Re^{\Box-V}_{0(2)} \right)
             +\frac{m^8}{\mu^9} \left( \frac{105}{8} u^2 \right)
             \right\} \, .  \label{5.9}
\eea

As earlier, we write 
\be \label{5.10}
\Delta L_{\omega} 
  =  \Delta L_{\omega}^{\sharp} 
     +\Delta L_{\omega}^{\flat} \, ,
\ee
where $\Delta L_{\omega}^{\sharp}$ is the part of the anomaly which 
dominates at high frequencies ($\omega\rightarrow\infty$), 
\bea
\Delta L_{\omega}^{\sharp}
  & = &  \frac{1}{4\pi} \left\{ \vphantom{ \frac{m^6}{\mu^7} }
             \frac{1}{3}\bar{\omega}^3 
             -\frac{1}{2}m\bar{\omega}^2
             +\bar{\omega} \left[ \frac{m^2}{2} 
                 -\frac{1}{2}\left( \frac{1}{6}R - V
                     - (\nabla\phi)^2 + \Delta_h\phi \right)
                 +\frac{5u}{2} - \Re^{\Box-V}_{0(2)} \right] 
             \right. \nonumber \\
  &   &      \hspace{0.33in} \mbox{}
             +\left[ \frac{m}{2}\left( \frac{1}{6}R - V
                 -(\nabla\phi)^2 + \Delta_h\phi \right)
                 -\frac{m^3}{3} \right]  \label{5.11} \\
  &   &      \hspace{0.33in} \left. \mbox{}
             +\frac{1}{\bar{\omega}} \left[           
                 \frac{3m^4}{8} 
                 -\frac{m^2}{4}\left( \frac{1}{6}R - V
                     -(\nabla\phi)^2 + \Delta_h\phi  
                     +3u - 2\Re^{\Box-V}_{0(2)} \right)
                 +\frac{1}{4}\Re^{\Box-V}_{2(0)} 
             \right] 
         \right\} \nonumber \, .
\eea
By subtracting this large-$\omega$ limit from the anomaly (\ref{5.9})
and making the inverse Fourier transform, we can construct an approximate 
effective Lagrangian for the four-dimensional theory:
\be\label{5.12}
L^F_{\ind{approx}} 
  =  \int_0^\infty \frac{d\bar{\omega}}{\pi} 
     \Delta L_{\omega}^{\flat}  \, .
\ee
Performing the $\omega$-integration (see Appendix \ref{mu_integrals}) gives
\bea
L^F_{\ind{approx}} 
  & = &  \frac{3m^4}{128\pi^2} 
         +\frac{m^2}{32\pi^2}\left[ 
             -\frac{1}{6}R+V+(\nabla\phi)^2-\Delta_h\phi 
             +u - 2\Re^{\Box-V}_{0(2)} 
         \right]  \nonumber \\
  &   &  \mbox{} 
         +\frac{1}{8\pi^2} \left[ \, \vphantom{\frac{1}{1}} 
             \Re^{\Box-V}_{1(2)}
             +4\Re^{\Box-V}_{0(4)}   
             -u\Re^{\Box-V}_{1(0)}
             -8u\Re^{\Box-V}_{0(2)}
             -8v +12u^2  
         \right]  \nonumber  \\
  &   &  \mbox{} 
         +\frac{1}{32\pi^2} \left[ \vphantom{\frac{1}{1}} \right.
             -\frac{m^4}{2} 
             +m^2 \left( 
                 \frac{1}{6}R-V-(\nabla\phi)^2+\Delta_h\phi 
                 +3u - 2\Re^{\Box-V}_{0(2)} 
             \right)
             \nonumber \\
  &   &      \hspace{0.55in} \mbox{}
             -\Re^{\Box-V}_{2(0)}   
         \left. \vphantom{\frac{1}{1}} \right] 
         \ln{ \left(\frac{m^2\e^{-4\phi}}{4\eta^2} \right) }
         ~. \label{5.13}
\eea
The effective action corresponding to this Lagrangian may be simplified 
considerably using integration by parts.  Substituting for the $u$, $v$, and 
$\Re^{\Box-V}_{n(k)}$ from Appendix~\ref{1+3_expansions} and neglecting surface
terms, one can show that the effective action for the important case 
$V=\xi\,^4\!R$ may be written as 
\bea
W^F_{\ind{approx}} 
  & = &  \int\! d^4\!x \sqrt{g} \left\{  
         -\frac{1}{64\pi^2} \ln{ \left(\frac{m^2\e^{-4\phi}}{4\eta^2} \right) }
         \left[ 
             m^4
             +\frac{1}{90}\left(
             \,^4\!R_{\alpha\beta\gamma\delta}\,^4\!R^{\alpha\beta\gamma\delta}
             -\,^4\!R_{\alpha\beta}\,^4\!R^{\alpha\beta}+\Box\,^4\!R \right)
         \right]
         \right. \nonumber \\[3mm]
  &   &  \hspace{-0.1in} \left. \mbox{}
         +\frac{3m^4}{128\pi^2} 
         -\frac{m^2(\nabla\phi)^2}{24\pi^2}
         +\frac{1}{360\pi^2}\left[ \,^4\!R_{\alpha\beta}\phi^\alpha\phi^\beta
             -\frac{3}{2}(\Box\phi)^2 - 4\Box\phi(\nabla\phi)^2 
             - 4(\nabla\phi)^4 \right]
         \right\}
         \nonumber \\[3mm]
  &   &  \mbox{} \hspace{-0.1in} 
         +\frac{\left(\xi-\frac{1}{6}\right)}{32\pi^2}
         \int\! d^4\!x \sqrt{g} \left\{
             m^2\,^4\!R
             -\frac{4}{3} \,^4\!R(\nabla\phi)^2
	     +\ln{ \left(\frac{m^2\e^{-4\phi}}{4\eta^2} \right) }
                 \left[ -m^2\,^4\!R + \frac{1}{6}\Box\,^4\!R  \right]
         \right\} \nonumber   \\[3mm]
  &   &  \mbox{} \hspace{-0.1in} 
         -\frac{\left(\xi-\frac{1}{6}\right)^2}{64\pi^2}
         \int\! d^4\!x \sqrt{g} \left\{
             \ln{ \left(\frac{m^2\e^{-4\phi}}{4\eta^2} \right) } \,(^4\!R)^2 
         \right\}
         \label{5.14}
\eea
Note that (\ref{5.14}) has been written in terms of four-dimensional 
quantities.  It may be shown \cite{FrSuZe:99} 
that the stress-energy tensor resulting from the
variation of (\ref{5.14}) in the special case of a 
static, spherically symmetric spacetime coincides 
with the analytic approximation 
of Anderson {\em et al} \cite{AnHiSa:95} for the zero-temperature case.  
Furthermore, in the massless, conformally-coupled limit (\ref{5.14})
coincides with the zero-temperature Killing approximation \cite{FrZe:87}.

\subsection{(2+2)-reduction}
\label{s5B}

As we already mentioned, the $(2+2)$-reduction is a special case
of the ``static''-spacetime reduction.  The calculation of the
dimensional-reduction anomaly is very similar to the
calculations of the previous subsection -- straightforward but
quite involved.  We do not reproduce the details of these calculations here
but simply give the final results.

The mode decomposition of the renormalized effective Lagrangian for the operator
$F = \Box-V-m^2$ has the form
\be \label{5.15}
L^F_{\ind{ren}} 
  =  \e^{2\phi} \int \frac{d{\bf p}}{(2\pi)^2} \left[ 
         L^{{\cal F}_{p}}_{\ind{ren}} + \Delta L_{p} \right]\, ,
\ee
where the dimensional-reduction anomaly $\Delta L_{p}$ is
\bea \label{5.16}
\Delta L_{p}=  
  & = &  \frac{1}{8\pi}  \left[ \;
         -\bar{p}^2 
         +\left( \frac{1}{6}R-V+\Delta\phi-(\nabla\phi)^2-m^2-\bar{p}^2 \right)
             \ln{ \left( \frac{m^2}{\mu^2} \right) }
         \right. \nonumber \\
  &   &  \hspace{0.33in} \mbox{} +\left( 12u - 4\Re^{\Box-V}_{0(2)} \right)
         +\frac{1}{\mu^2} \left( \Re^{\Box-V}_{2(0)} - 8m^2u 
             + 4m^2\Re^{\Box-V}_{0(2)} \right)
         \nonumber \\
  &   &  \hspace{0.33in} \mbox{} +\frac{m^2}{\mu^4} \left( - 4m^2u 
             + 4\Re^{\Box-V}_{1(2)} \right)
         +\frac{m^4}{\mu^6} \left( -8u\Re^{\Box-V}_{1(0)}  
             + 32\Re^{\Box-V}_{0(4)} \right)
         \nonumber \\
  &   &  \left. \hspace{0.31in} \mbox{} 
         +\frac{m^6}{\mu^8} \left( -96v -96u\Re^{\Box-V}_{0(2)} \right)
         +\frac{m^8}{\mu^{10}} \left( 192u^2 \right)  \; \right]  
         \, . 
\eea
Recall that $\bar{p}$ and $\mu$ are given by relation (\ref{4.35}).
By subtracting the high-frequency part and taking the inverse Fourier
transform we obtain a result for $W^F_{\ind{approx}}$ which 
identically coincides with (\ref{5.14}).

\section{Conclusion}\label{s6}
\setcounter{equation}0

In the presence of a continuous spacetime symmetry, when the field
equation can be solved by decomposition of the field into harmonics,
one can easily obtain similar mode-decomposed expressions for the field
fluctuations and the effective action.  Each of the terms in this
decomposition coincides with the corresponding object (the fluctuations
or the effective action) of the lower-dimensional theory obtained by
the reduction. We have demonstrated that because of the ultraviolet
divergences these decompositions have only formal meaning, as 
the renormalization violates the exact form of such
representations.  As a result, the expression for the renormalized
expectation value of the object in the ``physical'' spacetime can be
obtained by summing the contributions of corresponding lower-dimensional
quantities only if additional anomalous terms are added to each of the
modes.  We call this effect the {\em dimensional-reduction anomaly}.

In the general case, there can be different origins of the
dimensional-reduction anomaly.  In particular, such an anomaly may
arise when the lower-dimensional manifold which arises as the result of
the dimensional reduction has a non-trivial topology or has
boundaries.  We demonstrated how this kind of situation arises in the
simple examples of the reduction of four-dimensional flat spacetime in
spherical modes (Section~\ref{s3A}) and in Rindler time modes 
(Section~\ref{s3B}).
In addition to these ``global'' contributions to the dimensional-reduction
anomaly, there also exist ``local'' contributions.  The corresponding
anomalous terms are local invariants constructed from the curvature, 
the dilaton field, and their covariant derivatives.  Our main objective was
the study of such ``local'' dimensional-reduction anomalies.  We derived
the expressions for such anomalies for $(3+1)$-  and $(2+2)$- spacetime
reductions with a flat internal space. 
The remarkable fact that was discovered in this analysis is
that the calculated dimensional-reduction anomaly is closely related
to the {\em analytical approximation} in static spacetimes developed
in \cite{Ande:90,AnHiSa:95}.  We shall investigate this intriguing 
relationship further in a future publication \cite{FrSuZe:99}.

The dimensional-reduction anomaly discussed in the present paper
might have many interesting applications.  For example, it may  
be important for calculations of the contribution of individual 
modes with a fixed angular momentum to the stress-energy tensor and
the flux of Hawking radiation in the spacetime of a black hole.  In
order to obtain the contribution of each mode in 
the presence of the dimensional-reduction anomaly, the results of
two-dimensional calculations must be modified by terms produced by 
the anomaly.  The calculated dimensional-reduction anomaly for the 
effective action, or the corresponding result for the spherical reduction,
allows one to find directly the difference between the contribution
of a given mode to the stress-energy tensor, and the stress-energy
tensor of the corresponding lower-dimensional reduced theory.
This may also be important for the discussion of self-consistent
solutions which incorporate the back-reaction of the quantum fields.
We hope to return to the discussion of this and other
related topics in subsequent publications.


\bigskip

\vspace{12pt}
{\bf Acknowledgments}:\ \  This work was  partly supported  by  the
Natural Sciences and Engineering Research Council of Canada. Two of the
authors (V.F. and A.Z.) are grateful to the Killam Trust for its financial
support. 

\bigskip


\appendix
\section{Small-Distance Expansions for the (1+3) Reduction of
Static Space}
\label{1+3_expansions}

The line element for a static space may be written in the form
\be\label{A.1}
ds^2 = \e^{-4\phi(x)} dt^2 +  h_{ab}(x) dx^a dx^b ~.  
\ee
For this static metric, $h_{ab}$ is an induced 3-metric, and $n^\alpha =
\e^{2\phi(x)} \delta^\alpha_t$ is a unit vector normal to the surface
$t=\mbox{const}$. The extrinsic curvature $K_{ab}$ on $t=\mbox{const}$ 
surfaces vanishes. The nonvanishing Christoffel symbols are
\be\label{A.2}
{}^4\Gamma^a_{bc}[g] =  \Gamma^a_{bc}[h]\, ,  \hspace{0.5cm}
{}^4\Gamma^a_{00} = 2\e^{-4\phi} \phi^{;a}\, ,  \hspace{0.5cm}
{}^4\Gamma^0_{0a} =  -2 \phi_{;a} ~~. 
\ee
Because we will be using some quantities defined in terms of the full
four-dimensional metric $g$ and others in terms of 
the three-dimensional metric $h$, some
conventions on notation are in order.  Henceforth four-dimensional curvatures 
and covariant derivatives will be denoted by $^4\!R_{\cdots}$ and $()_{;a}$ 
respectively, while $\Box$ is understood to represent the d'Alembertian 
with respect to $g$.  
All other curvatures and covariant derivatives  
are understood to be calculated using the three-metric $h$. 
In particular, $\nabla$ and $()_{|a}$ are three-dimensional covariant
derivatives, and $\Delta = \Delta_h$ is the three-dimensional d'Alembertian.
For the dilaton $\phi$ we shall understand $\phi_a$, $\phi_{ab}$, etc. to
denote multiple three-dimensional covariant derivatives of $\phi$.

With these conventions we have, for example,
\be\label{A.3}
\Box\phi = \Delta\phi - 2(\nabla\phi)^2 ~.
\ee

It is convenient to define the following three-dimensional tensor
which occurs naturally in the $3+1$ reduction: 
\bea
T_{ab}  
  & = &  2 \left[ \phi_{ab} - 2 \phi_a \phi_b \right]  \, , \label{A.4} \\
T \; = \; T^a_a  
  & = &  2 \left[ \Delta \phi - 2 (\nabla \phi)^2 \right]  
         \; = \; 2 \Box \phi ~. \label{A.5}
\eea
In terms of $T_{ab}$ 
the only nonvanishing components of the four-dimensional 
curvatures are
\be \label{A.6}
^4\!R_{abcd} =  R_{abcd}\, ,  \hspace{0.5cm}
^4\!R_{0a0b} =  \e^{-4\phi} T_{ab}\, , 
\ee
\be \label{A.7}
^4\!R_{ab}  =  R_{ab} + T_{ab} \, ,   \hspace{0.5cm}
^4\!R_{00} =  \e^{-4\phi} T \, ,  \hspace{0.5cm}
^4\!R   =   R + 2T \, .
\ee

We shall also need  the following expressions for
$^4\!R_{\alpha\beta;\gamma}$,  
$^4\!R_{\alpha\beta;\gamma\delta}$,  
$^4\!R_{;\alpha}$, and
$^4\!R_{;\alpha\beta}$ : 
\bea
^4\!R_{ab;c}  & = &  \left(R_{ab}+T_{ab}\right)_{|c} \, ,  \nonumber  \\
^4\!R_{a0;0}  & = &  -2\e^{-4\phi} \left[ \left(R_{ab}+T_{ab}\right)\phi^b
                          -T \phi_a \right] \, , \nonumber \\
^4\!R_{00;c}  & = &  \e^{-4\phi} T_{|c} \, , \label{A.8}  \\
              &   &  \nonumber  \\
^4\!R_{ab;cd} & = &  \left(R_{ab}+T_{ab}\right)_{|cd} \, ,  \nonumber \\
^4\!R_{ab;00} & = &  -2\e^{-4\phi} \left[ 
                         \left(R_{ab}+T_{ab}\right)_{|c} \phi^c 
                         -4 \phi_a\phi_b T
                         +2 \phi_a \left(R_{bc}+T_{bc}\right) \phi^c 
                         +2 \phi_b \left(R_{ac}+T_{ac}\right) \phi^c 
		     \right] \, , \nonumber\\
^4\!R_{a0;c0} & = &  -2\e^{-4\phi} \left[ 
                         \left(R_{ab}+T_{ab}\right)_{|c} \phi^b 
                         +2 \phi^b \left( R_{ab} + T_{ab} \right) \phi_c
                         -2\phi_a\phi_c T - \phi_a T_{|c}  
                     \right] \, , \nonumber \\
^4\!R_{a0;0d} & = &  -2\e^{-4\phi} \left[
                         \left(R_{ab}+T_{ab}\right) \phi^b - T \phi_a
                     \right]_{|d} \, , \nonumber \\
^4\!R_{00;cd} & = &  \e^{-4\phi} T_{|cd} \, , \nonumber \\
^4\!R_{00;00} & = &  2\e^{-8\phi} \left[
                         4 \left( R_{ab} + T_{ab} \right) \phi^a \phi^b 
                         -4 T \phi^a \phi_a - T_{|a} \phi^a 
                     \right] \, , \label{A.9}  \\
              &   &  \nonumber \\
^4\!R_{;a}    & = &  \left( R+2T \right)_{|a} \, , \nonumber  \\
^4\!R_{;ab}   & = &  \left( R+2T \right)_{|ab}\, , \nonumber  \\ 
^4\!R_{;00}   & = &  - 2\e^{-4\phi} \left(  R+2T \right)_{|a} \phi^a \, .  
                     \label{A.10}
\eea

For the mode decomposition (Fourier time transform) we need to know 
the behaviour of the two-point functions $\sigma$ and 
$\Delta^{1/2} a_n^{\Box-V}$ for points $x^\alpha = (t,x)$ and 
$x'^\alpha = (t',x)$ where $\tau \equiv \e^{-2\phi} (t-t')$ is small. 
For $\sigma$ one can easily show that 
\bea 
2\sigma(t,x;t',x) 
   & = &  \tau^2 - \frac{1}{3}\phi^a\phi_a \tau^4 
          + \frac{1}{45} \left[ 8(\phi^a\phi_a)^2 - 3\phi^a\phi^b\phi_{ab}
          \right] \tau^6 + \cdots  \, , \nonumber    \\
\sigma_a(t,x;t',x)
   & = &  \mbox{} -\phi_a \tau^2 + \frac{1}{6} \left[ 
          \phi^b\phi_b\phi_a - \phi^b\phi_{ba} \right] \tau^4 + \cdots
	  \, , \label{A.11} \\
\sigma_t(t,x;t',x)
   & = &  \e^{-2\phi}\tau \left[ 1 - \frac{2}{3}\phi^a\phi_a \tau^2
          + \frac{1}{15} \left[ 8(\phi^a\phi_a)^2 - 3\phi^a\phi^b\phi_{ab} 
          \right] \tau^4 + \cdots  \right] \, .  \nonumber 
\eea

Combining the above expressions with the 
results of \cite{Chri:76,Chri:78} for 
small-$\sigma^\alpha$ expansions of the $\Delta^{1/2} a_n$,
it is easily shown that for the operator 
$\Box-V$, where the potential function $V$ is independent of $t$, the first
three $\Delta^{1/2} a_n^{\Box-V}$ are\footnote{
The Van Vleck determinant $\Delta$ (\ref{vanvleck}) is not to be confused 
with our notation for the three-dimensional covariant derivative.}
\bea
\Delta^{1/2} a_0^{\Box-V} 
     & = &  1 + \frac{1}{12} T \tau^2
            +\frac{1}{360} \left[ 6 \left( R_{ab}+T_{ab} \right) \phi^a\phi^b
                -16T\phi^a\phi_a + 6T_{|a}\phi^a 
                \right.\nonumber \\
     &   &      \left. \mbox{}   
                +\frac{5}{4}T^2 + T^{ab}T_{ab}
            \right] \tau^4 + O(\tau^6) \, , \nonumber \\
     &   &  \nonumber \\
\Delta^{1/2} a_1^{\Box-V} 
     & = &  \left(\frac{1}{6}R-V+\frac{1}{3}T\right)     
            +\left\{ 
                \frac{1}{2} \left(\frac{1}{6}R-V+\frac{1}{3}T\right)_{|a}\phi^a 
                +\left(\frac{1}{3}V_{|a} \phi^a - \frac{1}{12} V T \right) 
                \right. \nonumber \\ 
     &   &      \mbox{} 
                +\frac{1}{360} \left[
                    3\Delta T + 6T^2 + 6T^{ab}T_{ab} + 5RT + 2R^{ab}T_{ab} 
                    +24\left(R_{ab}+T_{ab}\right) \phi^a\phi^b 
                    \right. \nonumber  \\
     &   &          \left. \vphantom{\left(\frac{1}{3}\right)_{|a}} 
                    \mbox{} \left.  
                    -24 T\phi^a\phi_a - 18 R_{|a}\phi^a - 42T_{|a}\phi^a 
                \right]
            \right\} \tau^2 + O(\tau^4)\, , \label{A.12}   \\
     &   &  \nonumber \\
\Delta^{1/2} a_2^{\Box-V} 
     & = &  \frac{1}{2} \left(\frac{1}{6}R-V+\frac{1}{3}T\right)^2       
            +\frac{1}{30} \left[ \Delta (R+2T) - 2(R+2T)_{|a} \phi^a \right]   
            -\frac{1}{6} \left[ \Delta V - 2V_{|a} \phi^a \right]
	    \nonumber \\
     &   &  \mbox{} 
            +\frac{1}{180} \left[ 
                R_{abcd}R^{abcd} + 4T^{ab}T_{ab}
                -\left( R^{ab}+T^{ab} \right) \left( R_{ab}+T_{ab} \right)
                - T^2 
            \right] + O(\tau^2) \, . \nonumber 
\eea


\section{Small-Distance Expansions for the (2+2) Reduction}
\label{2+2_expansions}
\setcounter{equation}0

Consider a spacetime with the line element (\ref{4.29}).  In this case we
will be using some quantities defined in terms of the full
four-dimensional metric $g_{\mu\nu}$, and others in terms of the
two-dimensional  metric $h_{AB}$ $(A,B=2,3)$.  
In analogy to the $(3+1)$-splitting case,
four-dimensional curvatures and covariant derivatives will be denoted
by $^4\!R_{\cdots}$ and $()_{;a}$  respectively, while $\Box$ is
understood to represent the d'Alembertian  with respect to
$g_{\mu\nu}$.   All other curvatures and covariant derivatives are
understood to be calculated using the two-metric $h_{AB}$. In
particular, $\nabla$ and $()_{|A}$ are two-dimensional covariant
derivatives, and $\Delta = \Delta_h$ is the two-dimensional
d'Alembertian. For the dilaton $\phi$ we shall understand $\phi_A$,
$\phi_{AB}$, etc. to denote multiple two-dimensional covariant
derivatives of $\phi$. For example, with these conventions, the
four-dimensional d'Alembertian of a $y$-independent scalar $S$
decomposes to 
\be
\Box S = \Delta S - 2\nabla\phi\cdot\nabla S ~.
\ee
In particular, 
\be
\Box \phi  =  \Delta \phi - 2(\nabla\phi)^2  
=  -\frac{1}{2}\e^{2\phi}\Delta\e^{-2\phi}  ~.
\ee

For the given line element, the nonvanishing Christoffel symbols are
$(i,j =0,1)$
\bea   
\Gamma^A_{BC}  & = &  g^{AD} \Gamma_{DBC}  =  \frac{1}{2} h^{AD}
                      (h_{DB,C} + h_{CD,B}  - h_{BC,D} ) \, ,  \nonumber \\
\Gamma^A_{ij}  & = &  \phi^A {\rm e}^{-2\phi} \eta_{ij}  =  \phi^A g_{ij}
                      \, , \nonumber \\
\Gamma^i_{jA}  & = &  - \phi_A \delta^i_j  =  -\phi_A g^i_j  \, , \nonumber \\
\Gamma_{Aij}   & = &  -\Gamma_{ijA}  =  \phi_A {\rm e}^{-2\phi} \eta_{ij} 
                      =  \phi_A g_{ij} \,  .
\eea
Meanwhile, the only nonvanishing components of the four-dimensional 
curvatures are
\bea
^4\!R_{ABCD}  & = &  \frac{1}{2} R (g_{AC}g_{BD}-g_{AD}g_{BC}) \, , \nonumber \\
^4\!R_{AmBn}  & = &  g_{mn} [\phi_{AB} -\phi_A \phi_B]  \, , \nonumber \\
^4\!R_{ijkm}  & = &  - \phi_A \phi^A (g_{ik}g_{jm}-g_{im}g_{jk}) \, , \\
              &   &  \nonumber \\
^4\!R_{AB}    & = &  \frac{1}{2} R g_{AB} + 2 [\phi_{AB} - \phi_A\phi_B] 
                     \, , \nonumber\\
^4\!R_{mn}    & = &  g_{mn} [\phi^A_A - 2\phi^A\phi_A] \, ,  \\
              &   &  \nonumber \\
^4\!R         & = &  R + 4\phi^A_A - 6\phi^A\phi_A \,. 
\eea
For comparison, note that 
\ben
^2\!R_{ABCD}  =  \frac{1}{2} R (g_{AC}g_{BD}-g_{AD}g_{BC}) \, , 
\een
\be
^2\!R_{AB}  =  \frac{1}{2} R g_{Ab} \, ,\hspace{0.5cm}
^2\!R  =  R  \, . 
\ee
We shall also need the following components
of $^4\!R_{\alpha\beta;\gamma}$,  
$^4\!R_{\alpha\beta;\gamma\delta}$, $^4\!R_{;\alpha}$, and $^4\!R_{;\alpha\beta}$:
\bea
^4\!R_{AB;C}  & = &  \frac{1}{2} g_{AB} R_{,C} 
                     +2 [\phi_{AB} - \phi_{A}\phi_B]_{|C} \, ,\nonumber \\
^4\!R_{Am;n}  & = &  g_{mn} [ -\frac{1}{2} R \phi_A 
                         +\phi_A\Delta\phi - 2\phi^B\phi_{BA}] \, ,\nonumber \\
^4\!R_{mn;A}  & = &  g_{mn} [\Delta\phi-2(\nabla\phi)^2]_{|A}\, , \\
              &   &  \nonumber \\
^4\!R_{mn;AB} & = &  g_{mn} [\Delta\phi-2(\nabla\phi)^2]_{|AB} \, ,\nonumber \\ 
^4\!R_{ij;km} & = &  \left( g_{ik}g_{jm} + g_{im}g_{jk} \right) 
                     \left[ 
                         \frac{1}{2} R (\nabla\phi)^2 
                         -(\nabla\phi)^2 \Delta\phi 
                         +2\phi^A\phi^B\phi_{AB} 
                     \right] \nonumber \\
	      &   &  \mbox{} -\left( g_{ij}g_{km} \right) 
                     \left[ \Delta\phi-2(\nabla\phi)^2 \right]_{|A}\phi^A 
                     \, , \\
              &   &  \nonumber \\
^4\!R_{;A}    & = &  \left[ R+4\Delta\phi-6(\nabla\phi)^2 \right]_{,A}\, , \\
              &   &  \nonumber\\
^4\!R_{;AB}   & = &  \left[ R+4\Delta\phi-6(\nabla\phi)^2 \right]_{|AB} 
                     \, , \nonumber \\
^4\!R_{;mn}   & = &  -g_{mn} \left[ R+4\Delta\phi-6(\nabla\phi)^2 \right]_{|A}
                     \phi^A \, .        
\eea
Note again that operators and curvatures are with respect to the 
two-dimensional metric $h_{AB}$ unless explicitly labelled otherwise.  

We now write out the expansions of $\sigma(x,x;{\bf t}-{\bf t}')$ and 
the $\Delta^{1/2} a_n (x,x;{\bf t}-{\bf t}')$ for small separations.
Defining $\tau^i \equiv \e^{-\phi} (t^i-{t'}^i)$, we have
\bea
2\sigma(x,x;\tau) 
  & = &  \tau^2 - \frac{1}{12} (\nabla\phi)^2 \tau^4
         +\frac{1}{360} \left[ 4(\nabla\phi)^4 - 3\phi^A\phi^B\phi_{AB} 
         \right] \tau^6 + \cdots  \, , \nonumber \\
\sigma^i(x,x;\tau) 
  & = &  \e^\phi \tau^i \left[ 
             1 - \frac{1}{6} \phi_A\phi^A \tau^2 +
             \frac{1}{120} \left[ 
                 4(\phi_A\phi^A)^2 - 3\phi^A\phi^B\phi_{AB}  
             \right] \tau^4 + ... 
         \right] \, , \nonumber \\
\sigma^a(x,x;\tau)  
  & = &  -\frac{1}{2}\phi^A \tau^2 
         -\frac{1}{24} \phi_B \left(\phi^{BA}-2\phi^B\phi^A\right) \tau^4  \\
  &   &  \mbox{}
         +\frac{1}{720} \left[ \vphantom{\frac{1}{1}}  
             -12(\phi^B\phi_B)^2 \phi^A 
             +8\phi^B\phi_B\phi_C\phi^{CA} + 9\phi^A\phi^B\phi^C\phi_{BC}
             \right. \nonumber \\
  &   &      \left.  \mbox{} 
             -3\phi^B\phi_{BC}\phi^{CA} - \frac{3}{2}\phi_B\phi_C\phi^{BCA}
         \right] \tau^6 + ...  \, . \nonumber
\eea
Combining these expressions with the results of \cite{Chri:76,Chri:78},
it is easily shown that for the operator 
$\Box-V$, where the potential function $V$ is independent of $t^i$, the first
three $\Delta^{1/2} a_n^{\Box-V}$ are
\bea
\Delta^{1/2} a_0^{\Box-V} 
   & = &  1 + \frac{1}{12} \left[ \Delta\phi - 2(\nabla\phi)^2 \right] \tau^2 
          +\frac{1}{1440} \left[ 
              3R(\nabla \phi)^2 + 48(\nabla \phi)^4 
              -36(\nabla\phi)^2\Delta\phi  
              \right.  \nonumber\\
   &   &      \left. \mbox{} 
              -44\phi^A\phi^B\phi_{AB} + 5(\Delta\phi)^2 
              +4\phi^{AB}\phi_{AB} + 12\phi^A(\Delta\phi)_A 
          \right] \tau^4 + O(\tau^6)   \, , \nonumber   \\
   &   &  \nonumber \\
\Delta^{1/2} a_1^{\Box-V} 
   & = &  \left[ \frac{1}{6}R-V+\frac{2}{3}\phi^A_A-\phi^A\phi_A \right]     
          \nonumber  \\
   &   &  \mbox{} 
          +\frac{1}{2} \left[ 
              -\frac{1}{6} V_A \phi^A 
              -\frac{1}{6} V [\Delta\phi - 2(\nabla \phi)^2]
              +\frac{1}{30} R_{,A} \phi^A 
              +R[\frac{1}{30}\Delta\phi - \frac{2}{45}(\nabla \phi)^2] 
              \right. \nonumber \\
   &   &      \mbox{} 
              +\frac{1}{180} \left( 
                  60(\nabla\phi)^4 
                  -62(\nabla\phi)^2\Delta\phi - 52\phi^A\phi^B\phi_{AB} 
                  +16(\Delta\phi)^2 - 4\phi^{AB}\phi_{AB} 
                  \right. \nonumber\\
   &   &          \left. \left.  \mbox{} 
                  +18\phi^A(\Delta\phi)_A
                  -12 \phi^A \Delta(\phi_A) + 3\Delta^2\phi
              \right) 
          \right] \tau^2 + O(\tau^4) \, , \\
   &   &  \nonumber \\
\Delta^{1/2} a_2^{\Box-V} 
   & = &  \frac{1}{2} \left[
              \frac{1}{6}R - V + \frac{2}{3}\phi^A_A - \phi^A\phi_A 
          \right]^2  
          - \frac{1}{6}\Delta V + \frac{1}{3}V_{,A}\phi^A 
          + \frac{1}{30}\Delta R - \frac{1}{15}R_{,A}\phi^A  
          \nonumber \\
   &   &  \mbox{} 
          +\frac{1}{180} \left[ 
              \frac{1}{2} R^2 
              -2R[\Delta\phi - (\nabla\phi)^2] + 8(\nabla\phi)^2\Delta\phi  
              +136\phi^A\phi^B\phi_{AB} - 2(\Delta\phi)^2 
              \right. \nonumber \\ 
   &   &      \left. \vphantom{\frac{3}{2}} \mbox{} 
              -68\phi^{AB}\phi_{AB} - 72\phi^A\Delta(\phi_A)
              -48\phi^A(\Delta\phi)_A + 24\Delta^2\phi  
          \right] + O(\tau^2) \, . \nonumber   
\eea


\section{Useful Formulae}
\label{C}
\setcounter{equation}0

\subsection{The Modified Bessel Function $K_\nu$}
\label{K_formulae}	

The modified Bessel functions $K_\nu(z)$ may be defined via the integral
\be\label{3int1}
\int_0^\infty dx x^{-1-\nu} \exp{ \left\{ -x -\frac{z^2}{4x} \right\} }
  =  2 \left( \frac{2}{z} \right)^\nu K_\nu(z)
\ee 
It may be shown that $K_{-\nu}(z) =  K_\nu(z)$.  Furthermore, for $\nu>0$ 
the $K_\nu(z)$ obey the differential relation
\be\label{3int2}
\left( - \frac{1}{z} \frac{d}{dz} \right)^n z^\nu K_\nu(z) 
  =  z^{\nu-n} K_{\nu-n}(z)
\ee
In particular, for $z=m\sqrt{\tau^2+\epsilon^2}$ one can easily show that 
\be\label{3int3}
\frac{1}{z^n} K_{n}(z)  =  
\left( - \frac{1}{m^2\epsilon} \frac{d}{d\epsilon} \right)^n K_0(z) 
\ee
Combining (\ref{3int3}) with integral (6.677) of \cite{GrRy:94},
\be\label{3int4}
\int_{-\infty}^\infty d\tau~
\cos(\bar{\omega}\tau)~K_0(m\sqrt{\epsilon^2+\tau^2})
  =  {\pi\over\sqrt{m^2+\bar{\omega}^2}}
     \exp(-\epsilon\sqrt{m^2+\bar{\omega}^2}) ~,
\ee
allows us to evaluate the $(1+3)$-splitting 
Fourier transforms of Sections~\ref{s4A}, \ref{s5A} as follows: 
\bea
\int_{-\infty}^\infty d\tau~
\cos(\bar{\omega}\tau)~\tau^{2k}
~\frac{1}{(m\sqrt{\tau^2+\epsilon^2})^n}K_n(m\sqrt{\epsilon^2+\tau^2}) 
  & = &   \nonumber \\
  &   &  \hspace{-3.3in} = (-1)^k \frac{d^{(2k)}}{d\bar{\omega}^{(2k)}} 
         \left( - \frac{1}{m^2\epsilon} \frac{d}{d\epsilon} \right)^n
         {\pi\over\sqrt{m^2+\bar{\omega}^2}}
         \exp(-\epsilon\sqrt{m^2+\bar{\omega}^2})
         \label{3int5}
\eea
For convenience, we have used the 
notation $\bar{\omega}=\e^{2\phi}\omega$ introduced in 
Section~\ref{s4A}.


\subsection{Integrals of $\mathbf{\mu^n}$ for the (1+3) Reduction}
\label{mu_integrals}    
 
For $\mu = \sqrt{m^2 + x^2}$ it is easily shown that for large $\bar{\omega}$ 
\bea
\int_{0}^{\bar{\omega}} dx \mu^3 
  & = &  \frac{1}{4}\bar{\omega}^4 + \frac{3}{4} m^2 \bar{\omega}^2
         + \frac{9}{32} m^4 + \frac{3}{8}m^4 \ln{ \frac{2\bar{\omega}}{m} } 
         \nonumber ~, \\
\int_{0}^{\bar{\omega}} dx \mu
  & = &  \frac{1}{2} \bar{\omega}^2
         + \frac{1}{4} m^2 + \frac{1}{2}m^2 \ln{ \frac{2\bar{\omega}}{m} } 
         \label{C.2.1} ~, \\
\int_{0}^{\bar{\omega}} dx \frac{1}{\mu}
  & = &  \ln{ \frac{2\bar{\omega}}{m} }  \nonumber  ~.
\eea
In addition, for $n \ge 1$,  
\be\label{C.2.2}
\int_0^\infty dx \mu^{-(2n+1)} 
  =  {1\over m^{2n}}{2^{n-1}\, (n-1)!\over (2n-1)!!}\, .
\ee
See, for example, (2.271) of \cite{GrRy:94}.
These results are sufficient to perform the sum over modes in
(\ref{4.27}, \ref{5.12}).

\subsection{Formulae for the (2+2) Reduction}
\label{2+2_transforms}

The Fourier transforms of Sections \ref{s4B}, \ref{s5B} 
were computed before performing 
the $s$-integration by expanding all $z$-dependent quantities for small
curvatures and using 
\be\label{C.3.1}
\int_{-\infty}^\infty d^2z \exp{ \left\{ -\frac{1}{4s}z^2+i\bar{p}z \right\} }
 z^{2n} = (4\pi s) \e^{-\bar{p}^2s} (4s)^n I_n  ~,
\ee
where $\bar{p} = \e^\phi p$ and 
\bea
I_0  & = &  1    \nonumber \\
I_1  & = &  1 - \bar{p}^2 s  \nonumber  \\
I_2  & = &  2 - 4\bar{p}^2 s + \bar{p}^4 s^2   \label{C.3.2}     \\
I_3  & = &  6 - 18\bar{p}^2 s + 9\bar{p}^4 s^2 - \bar{p}^6 s^3  \nonumber    \\
I_4  & = &  24 - 96\bar{p}^2 s + 72\bar{p}^4 s^2 - 16\bar{p}^6 s^3
            + \bar{p}^8 s^4  ~. \nonumber 
\eea       
For the summation over modes referred to in Sections~\ref{s4B}, \ref{s5B}
one may use the integrals 
\bea
\int_0^{\bar{p}} dx \, x \, \ln{\frac{\mu^2}{m^2}}
  & = &  \frac{1}{2} (\bar{p}^2+m^2) \ln{ \frac{\bar{p}^2+m^2}{m^2} } 
         - \frac{1}{2} \bar{p}^2 \nonumber  ~,  \\
\int_0^{\bar{p}} dx \, x \, \mu^{-2}
  & = &  \frac{1}{2} \ln{ \frac{\bar{p}^2+m^2}{m^2} } ~, 
\eea
and for $n > 1$,
\be
\int_0^\infty dx \, x \, \mu^{-2n} = \frac{1}{2(n-1)m^{2(n-1)}} ~,
\ee
where $\mu = \sqrt{m^2+x^2}$.



\end{document}